\documentclass[final,5p,times,twocolumn,numbers,compress]{elsarticle}
\usepackage[T1]{fontenc}
\usepackage{hyperref}
\usepackage{graphicx}
\usepackage{dcolumn}
\usepackage{bm}
\usepackage{appendix}
\usepackage{float}
\usepackage[normalem]{ulem} 
\usepackage{amsmath}
\usepackage{booktabs}
\usepackage{xcolor}
\usepackage[percent]{overpic}
\newcommand{\Dipper}{\textsc{McDipper}}
\newcommand{\trento}{\textsc{TrENTo}}
\begin{document}
	
	\begin{frontmatter}
		
		\title{Fundamental geometric limitations on disentangling nuclear-surface properties in relativistic heavy ion collisions}

		\author[first]{Hadi Mehrabpour}
		\ead{mehrabpour@fudan.edu.cn}
		\author[second]{Behnaz Behzadmoghaddam}
		\ead{b.behzadmoghaddam@ipm.ir}
		\author[second]{S.~M.~A.~Tabatabaee Mehr}
		\ead{tabatabaee@ipm.ir}
		\author[third]{Oscar Garcia-Montero}
		\ead{oscarjesus.garcia@usc.es}
		\author[first]{Li Yan}
		\ead{cliyan@fudan.edu.cn}  
		\address[first]{Institute of Modern Physics and Key Laboratory of Nuclear Physics and Ion-beam Application (MOE), Fudan University, Shanghai 200433, China}
		\address[second]{School of Particles and Accelerators, Institute for Research in Fundamental Sciences (IPM), P.O. Box 19395-5531, Tehran, Iran}
		\address[third]{Instituto Galego de Física de Altas Enerxías IGFAE, Universidade de Santiago de Compostela, E-15782 Galicia, Spain}
		
	\begin{abstract}
		The extraction of the nuclear surface diffuseness from relativistic heavy ion collisions is limited by the intertwined responses of geometry-driven observables to surface diffuseness and intrinsic nuclear
		deformation. We investigate this limitation using event-by-event Monte Carlo Glauber simulations, focusing on the sensitivity of multiparticle
		correlations to the Woods--Saxon surface diffuseness $a_0$ in intrinsically deformed nuclei. We systematically examine the local correlations between
		$a_0$ and quadrupole and octupole deformation parameters, $\beta_2$ and $\beta_3$, and determine how these correlations affect the
		ability of different observables to constrain $a_0$. We find that observables dominated by elliptic geometry exhibit a strong response to quadrupole deformation, leading to a local $a_0$--$\beta_2$ degeneracy that
		substantially limits their sensitivity to the nuclear surface diffuseness. Triangular correlations provide a more independent response to the nuclear
		surface and therefore retain additional information on $a_0$, although their sensitivity can also be modified by intrinsic deformation. Extending the analysis to simultaneous quadrupole and octupole deformation shows that the local degeneracy and least-constrained directions depend on the nuclear configuration, demonstrating that the limitation on extracting
		$a_0$ is not described by a single global parameter correlation. We quantify these effects using multidimensional response maps, local
		sensitivity and information-geometric measures, and a Bayesian analysis of the resulting parameter constraints. The results clarify the fundamental limitations imposed by intrinsic multipole deformation on the
		determination of nuclear surface diffuseness from relativistic heavy ion collisions and identify multiparticle correlations that provide more independent information on $a_0$.
	\end{abstract}
	\end{frontmatter}

\section{Introduction}

The Woods--Saxon (WS) parameterization provides a standard description of
nuclear density distributions in terms of the half-density radius, surface
diffuseness $a_0$, and intrinsic deformation parameters. It reproduces
ground state densities obtained from nuclear density functional theory
(DFT)~\cite{Bender2003,Schunck2019} and provides a convenient connection
between nuclear geometry and the underlying many-body structure
~\cite{Fricke:1995zz}. While charge densities are known with high precision
from elastic electron scattering~\cite{DeVries1987}, the neutron density
profile and, in particular, the properties of the nuclear surface remain
less constrained because neutrons cannot be directly accessed through
electromagnetic probes~\cite{Thiel:2019tkm}. Surface diffuseness and
intrinsic deformations are therefore commonly inferred through
model-dependent analyses of hadronic scattering, antiprotonic atoms,
parity-violating electron scattering, and nuclear spectroscopy
~\cite{Brown:2000pd,PREX:2021umo,CREX:2022kgg}. Disentangling these
geometric parameters remains a longstanding challenge
~\cite{Pihan:2025pep,Zhao:2026zno}, since different WS parameters modify
the nuclear geometry simultaneously and can generate similar responses
in geometric observables~\cite{Ryssens:2023fkv}. In particular, extracting
$a_0$ in a deformed nucleus requires separating the response to the nuclear
surface from that induced by intrinsic deformation.

Ultrarelativistic nuclear collisions address this challenge by converting initial geometric fluctuations into collective final-state particle correlations through hydrodynamic medium response~\cite{Heinz2013,Alver2010,Mehrabpour:2026lhj,Mehrabpour:2026yuc,Mehrabpour:2023ign,Mehrabpour:2025rzt,Parida:2026uld,Giacalone:2026fat,Giacalone:2017dud,Rybczynski:2019adt,Summerfield:2021oex,Zhang:2021kxj,Xu:2021uar,Nijs:2021kvn,Zhao:2022uhl,Samanta:2023qem,Giacalone:2023cet,Fortier:2023xxy,Xu:2024bdh,Zhang:2024vkh,Zhao:2024feh,Fortier:2024yxs,Giacalone:2024ixe,Mantysaari:2024uwn,Lu:2025cni,Li:2025vdp,Liu:2025zsi,Li:2025hae,Zhang:2025hvi,TabatabaeeMehr:2024lgu,Taghavi:2025ddm,Li:2026igf}, directly bridging initial many-body structure to experimental observables~\cite{Bofos:2026nmg,Mehrabpour:2025ogw,Liu:2025uks,Giacalone:2023hwk,Blaizot:2025bfu,Duguet:2025qxi,Blaizot:2025scr,Blaizot:2026yvx}. This connection between nuclear structure and final-state
observables has already led to several important developments. Jia
\textit{et al.} demonstrated that isobar collisions provide simultaneous
sensitivity to neutron skin thickness and nuclear deformation, allowing
the two effects to be disentangled through collective flow measurements
~\cite{Jia2022b}. Giacalone \textit{et al.} subsequently constrained the
neutron skin thickness of $^{208}\mathrm{Pb}$ using ultrarelativistic
heavy ion collisions at the LHC~\cite{Giacalone2023}. More recently,
Pihan \textit{et al.} extracted neutron-skin information from conserved-charge observables in $p+\mathrm{A}$ collisions while allowing the neutron
surface diffuseness to vary~\cite{Pihan:2025pep}, and Vitsos
\textit{et al.} extended related studies to the neutron-rich nucleus
$^{48}\mathrm{Ca}$~\cite{Vitsos:2025jzt}. Together, these developments
establish relativistic heavy ion collisions as an emerging probe of
nuclear density distributions complementary to traditional low-energy
measurements~\cite{Sun:2026yrr,Li:2026yzw}.

A key limitation remains: the response of a given
multiparticle correlation need not be specific to the nuclear surface.
Intrinsic deformation can dominate the same geometric observables and
thereby obscure their dependence on $a_0$~\cite{Giacalone:2025vxa}. This issue is particularly
relevant for deformed nuclei with simultaneous quadrupole and octupole
deformations, characterized by $\beta_2$ and $\beta_3$, respectively.
The resulting correlations between $a_0$ and $\beta_{\lambda}$,
$\lambda=2,3$, can make different nuclear configurations locally
indistinguishable through individual observables. The relevant question is
therefore not simply whether heavy ion observables respond to $a_0$, but
whether that response remains independently identifiable once intrinsic
deformation is allowed to vary.

In this work, we address this question using event-by-event Monte Carlo
simulations of $^{238}\mathrm{U}+^{238}\mathrm{U}$~\cite{STAR:2025elk} and
$^{20}\mathrm{Ne}+^{20}\mathrm{Ne}$~\cite{Giacalone:2024luz,Raman2001}
collisions. We systematically vary the WS surface diffuseness together
with quadrupole and octupole deformations and quantify the resulting
response of two- and three-particle correlation observables. We find that
correlations associated with elliptic geometry are predominantly governed
by quadrupole deformation, substantially limiting their independent
sensitivity to $a_0$, while triangular correlations retain a more distinct
response to the nuclear surface. We further show that the resulting
deformation--diffuseness degeneracy is local and depends on the underlying
nuclear configuration, with the least-constrained directions in parameter
space changing when $\beta_2$ and $\beta_3$ are varied simultaneously.
The qualitative response structure remains robust against variations of
the nucleon width and short-range nucleon correlations in the
\trento~model~\cite{Moreland:2014oya}, as well as in the \Dipper~\cite{Garcia-Montero:2023gex,Garcia-Montero:2024jev,Garcia-Montero:2025bpn}, an initial-state model based on the $k_T$-factorization of Color Glass Condensate (CGC) effective description of QCD 
~\cite{Garcia-Montero:2025hys} with a
nucleon profile constrained by DIS data~\cite{Rose:2014fba,Ryu:2015vwa,
	Ryu:2017qzn}. Finally, we quantify the resulting parameter constraints
within a Bayesian framework and use local information-geometric measures
to identify the combinations of nuclear parameters that remain weakly
constrained. These results establish the limitations imposed by intrinsic
quadrupole and octupole deformation on extracting the nuclear surface
diffuseness from relativistic heavy ion collisions.

\section{Models and sensitivity analysis}\label{sec:models}

We generate the initial nuclear geometry event-by-event with the Monte Carlo
Glauber model~\cite{Moreland:2014oya} implemented in \trento, with selected comparisons to the
CGC-based \Dipper\ framework. The nuclear density is described by the
deformed WS form~\cite{Miller:2007ri,PhysRev.95.577},
\begin{align}
	\rho(r,\theta,\phi)&=
	\rho_0\left[1+\exp\Big((r-R(\theta,\phi))/a_0\Big)\right]^{-1},
	\\\
	R(\theta,\phi)&=R_0\Big[1+\sum_{\ell,m}\beta_{\ell m}
	Y_\ell^m(\theta,\phi)\Big].
	\label{eq:WS}
\end{align}
Here, $a_0$ is the surface diffuseness, $R_0$ is the half-density radius,
and $\beta_{\ell m}$ characterize the intrinsic deformation~\cite{BohrMottelson1975,RingSchuck1980,Li2024Rn224,Robledo2025History}. We fix $R_0$ ($R_{0}^{U}=6.81$ fm \cite{STAR:2024wgy} and $R_{0}^{Ne}=2.801$ fm \cite{ANGELI201369})
to its experimentally established value and vary $a_0$, $\beta_2$, and
$\beta_3$ over the parameter space considered below. Since the orientation
of the nucleus is randomized event by event, the sign of the octupole
deformation is physically equivalent to an inversion of the pear-shaped
density; we therefore restrict the scan to $\beta_3\geq0$~\cite{Zhang:2025zrm}. Variations of
the nucleon width and minimum inter-nucleon separation in \trento\ are used
to assess model dependence, together with comparisons to \Dipper.

The fluctuating initial geometry is characterized by multiparticle
correlations that can be related to final-state observables through the
collective response of the produced medium in the linear response of hydrodynamics,
$v_n\simeq\kappa_n\varepsilon_n$. Transverse-size fluctuations are
related to mean-$p_T$ fluctuations through
$\delta[p_T]/[p_T]\propto\delta d_\perp/d_\perp$,
where $d_\perp$ characterizes the inverse transverse size of the overlap
region~\cite{Jia:2021qyu}. We therefore consider the two-particle
correlation measures $\varepsilon_2\{2\}$, $\varepsilon_3\{2\}$, and
$c_d\{2\}=\langle(\delta d_\perp/d_\perp)^2\rangle$, together with the
three-particle flow--transverse-size correlations
\begin{equation}
	\mathrm{cov}(n)=
	\left\langle\varepsilon_n^2\frac{\delta d_\perp}{d_\perp}\right\rangle
	-\left\langle\varepsilon_n^2\right\rangle
	\left\langle\frac{\delta d_\perp}{d_\perp}\right\rangle,
	\qquad n=2,3.
	\label{eq:covariance}
\end{equation}
These quantities provide different projections of the fluctuating
many-body nuclear geometry and are related to experimentally accessible
flow and flow--mean-$p_T$ correlations
~\cite{Niemi2013,Teaney:2013dta,Niemi:2012aj,Giacalone:2020dln,
	Jia:2021tzt,Bozek:2016yoj,Bozek:2020drh,Schenke:2020uqq}.

We quantify the local dependence of a correlation $X$ on the nuclear parameters through the sensitivity vector. Using the scaled WS parameters introduced in \ref{app:bayesian}, we define
\begin{equation}
	\mathbf S_X=
	\left(
	\partial X/\partial x_{a_0},
	\partial X/\partial x_{\beta_2},
	\partial X/\partial x_{\beta_3}
	\right),
\end{equation}
where the $x$ variables are related linearly to the physical parameters
$(a_0,\beta_2,\beta_3)$. Its magnitude measures the local response, while
its direction specifies the corresponding direction in the chosen
parameter coordinates. The local complementarity of two correlations is
characterized by the normalized scalar product
\begin{equation}
	\cos\psi=
	\mathbf S_{X_i}\!\cdot\!\mathbf S_{X_j}/
	(|\mathbf S_{X_i}|\,|\mathbf S_{X_j}|),
	\label{eq:cosphi}
\end{equation}
with values near $\pm1$ corresponding to parallel or antiparallel
responses and values near zero to orthogonal responses. The sensitivity
vectors are assembled into the Jacobian,
\begin{equation}
	J_{ij}=
	\left.
	\partial X_i/\partial x_j
	\right|_{\mathbf{x}_{\rm ref}}, 
\end{equation}
whose singular-value decomposition determines the principal response
directions and the least-constrained local parameter combination. We additionally
construct the uncertainty-weighted Fisher-information matrix to quantify the combined information carried by the correlations while accounting for
statistical and model uncertainties. These tools provide the basis for the
local-degeneracy and information-geometry analysis presented in
Sec.~\ref{sec:results}.

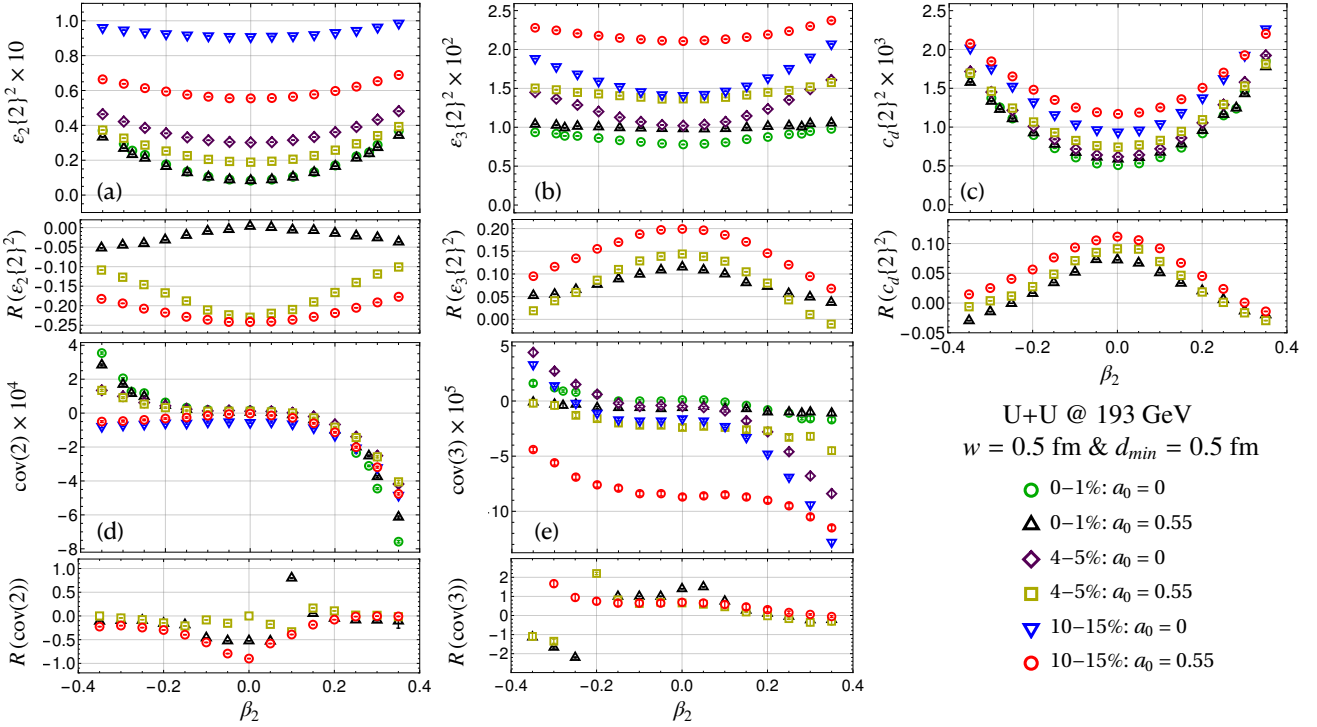
\begin{figure*}[t!]
	\includegraphics[scale=.26]{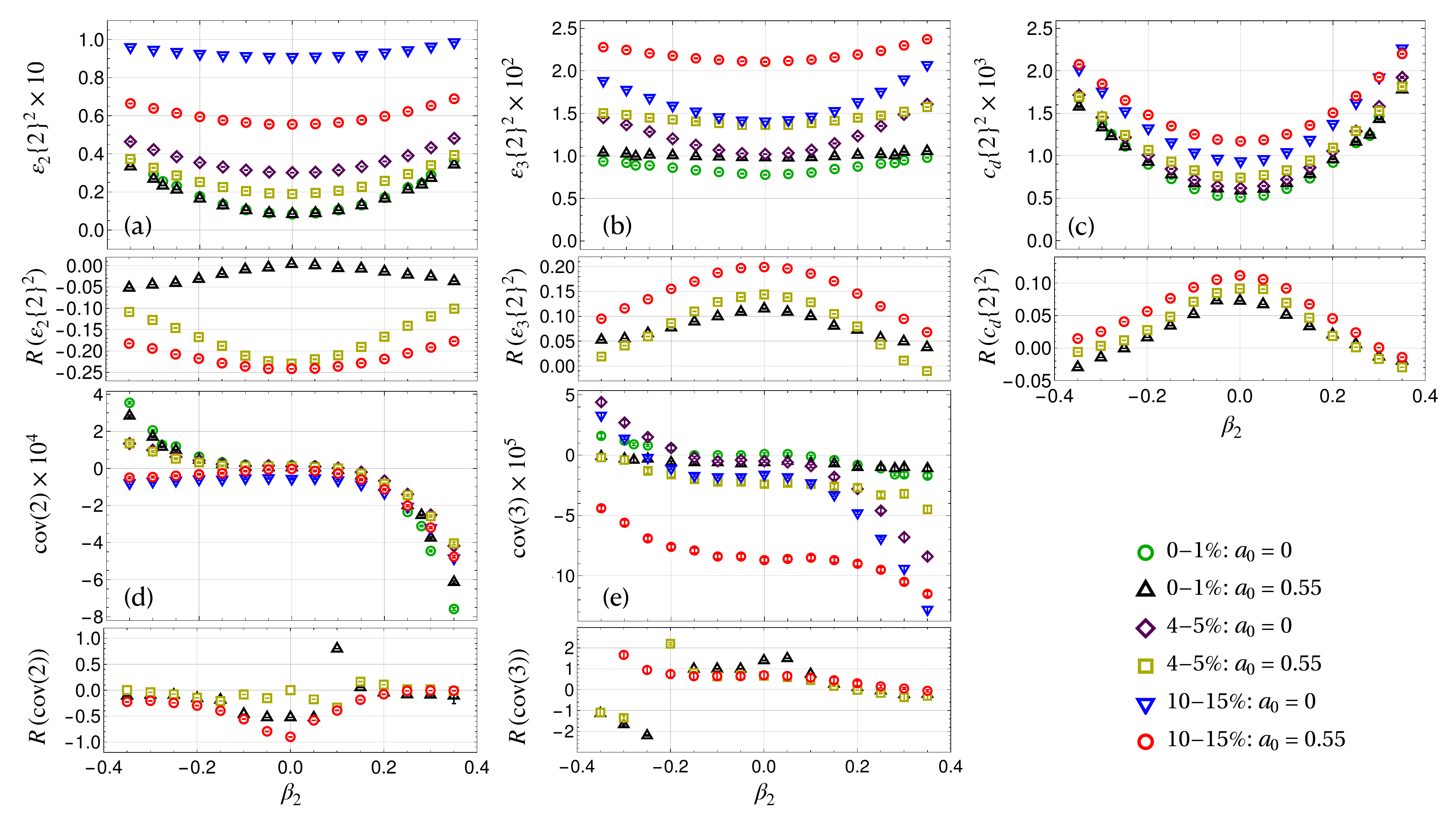}
	\begin{picture}(0,0)
		\put(-120,117){{\fontsize{10}{10}\selectfont \textcolor{black}{U+U @ 193 GeV}}}
		\put(-135,105){{\fontsize{10}{10}\selectfont \textcolor{black}{$w = 0.5$ fm \& $d_{min}=0.5$ fm}}}
	\end{picture}	
	\caption{
		Initial-state observables as a function of the quadrupole deformation
		$\beta_2$ for $w=d_{\min}=0.5$~fm. The upper subplots show
		$\varepsilon_2\{2\}^2$, $\varepsilon_3\{2\}^2$, $c_d\{2\}$,
		$\mathrm{cov}(2)$, and $\mathrm{cov}(3)$ for two representative values
		of the surface diffuseness, $a_0=0$ and $0.55$~fm, in the
		$0$--$1\%$, $4$--$5\%$, and $10$--$15\%$ centrality classes.
		The lower subplots show the corresponding normalized response of Eq.\ref{eq:R_a0} to changes in surface diffuseness for the $0$--$1\%$ (black triangles), $4$--$5\%$ (yellow rectangles), and $10$--$15\%$ (red circles) centrality classes, following the same format as the upper subplots.
		The results demonstrate that the sensitivity to $a_0$ depends strongly
		on both the observable and $\beta_2$, and varies with centrality.
		Consequently, the effect of surface diffuseness cannot in general be
		isolated independently of the intrinsic quadrupole deformation.}
	\label{fig:response_a0}
\end{figure*}
Finally, we use the calculated correlations in a Bayesian response-model
analysis. Each correlation is represented by a smooth interpolating
response over the scanned WS parameter space, and the likelihood includes
both Monte Carlo uncertainties and a model-discrepancy contribution
estimated from the response-model residuals. The posterior and
posterior-predictive distributions are then evaluated for the relevant
nuclear parameters. The response model, discrepancy treatment, likelihood,
and posterior construction are detailed in \ref{app:bayesian}.

\section{Results and discussion}\label{sec:results}

The multiparticle correlations generated by the fluctuating nuclear geometry provide different projections of the underlying WS density. A central question is therefore whether these correlations retain independent information about the surface diffuseness $a_0$, or whether
this information is obscured by the simultaneous presence of intrinsic quadrupole deformation $\beta_2$.

\subsection{Sensitivity to surface diffuseness and quadrupole deformation:} Figure~\ref{fig:response_a0} illustrates
this interplay for U+U collisions with
$w=d_{\min}=0.5~\mathrm{fm}$. The upper panels show the two-particle
correlation measures $\varepsilon_2\{2\}$, $\varepsilon_3\{2\}$, and
$c_d\{2\}$, together with the three-particle correlations
$\mathrm{cov}(2)$ and $\mathrm{cov}(3)$, as functions of $\beta_2$ for
two representative values of the surface diffuseness,
$a_0=0$ and $a_0=0.55~\mathrm{fm}$. The corresponding lower panels show
the normalized response of each observable $X$ to the change in surface diffuseness,
\begin{equation}
	R(X)
	=
	\frac{ X(a_0=0.55)-X(a_0=0)}
	{X(a_0=0.55)+X(a_0=0)} ,
	\label{eq:R_a0}
\end{equation}
which isolates the relative change of each correlation measure while
retaining its dependence on $\beta_2$ and collision centrality.

The two-particle correlations exhibit markedly different
responses to the nuclear surface. For $\varepsilon_2\{2\}$, the relative
response to $a_0$ is generally small in the most central collisions,
with $|R|\lesssim0.05$ around $\beta_2=0$, while it becomes substantially
larger at finite impact parameter, reaching approximately
$R\simeq-(0.2-0.25)$ in the $10$--$15\%$ centrality interval. This weak
response is a consequence of the strong sensitivity of the elliptic
correlation to $\beta_2$, which can dominate over the change induced by
the radial surface profile. In contrast, $\varepsilon_3\{2\}$ exhibits
a positive response to increasing $a_0$, reaching approximately
$R\simeq0.1$ in the $0$--$1\%$ interval and about $R\simeq0.2$ around
$\beta_2\simeq0$ in the $10$--$15\%$ interval. The transverse-size
correlation $c_d\{2\}$ also increases with $a_0$, with a response that
becomes more pronounced away from the most central collisions and
reaches values of order $R\sim0.1$. Thus, even among the two-particle
correlations, the relative information carried by the nuclear surface
is strongly correlation-dependent.

The three-particle correlations provide a complementary view of this
interplay. In particular, $\mathrm{cov}(2)$ exhibits a strong dependence
on $\beta_2$ and can change sign as the deformation is varied. As a
result, its normalized response can become very large when the
denominator of Eq.~(\ref{eq:R_a0}) becomes small; in the
$10$--$15\%$ interval, values approaching $R\simeq-1$ occur around
$\beta_2\simeq0$. Similarly, $\mathrm{cov}(3)$ shows a pronounced
dependence on both $\beta_2$ and centrality, with normalized responses
of order unity and, in some regions, greater than unity. These large
values of $R(X)$ should not be interpreted directly as a larger
absolute sensitivity to $a_0$, since the ratio depends on both the
magnitude and sign of the underlying correlation. Nevertheless, the
strong variation demonstrates that the three-particle correlations
retain substantial information about changes in the nuclear density
profile and, importantly, probe this information through combinations
that differ from those entering the two-particle correlations.

The dependence on $\beta_2$ is essential for interpreting these responses.
At fixed centrality, the separation between the correlation curves
corresponding to $a_0=0$ and $a_0=0.55~\mathrm{fm}$ varies continuously
with $\beta_2$, and several correlations exhibit rapid changes or even
sign reversals. Consequently, the magnitude of a measured multiparticle
correlation cannot, in general, be mapped onto a unique value of the
surface diffuseness without simultaneously accounting for the intrinsic
quadrupole deformation. Conversely, variations of $\beta_2$ can
partially compensate for changes in $a_0$, producing locally similar
correlation signals from different nuclear density profiles. This
deformation--diffuseness degeneracy is precisely the issue that motivates
the local sensitivity analysis developed below.
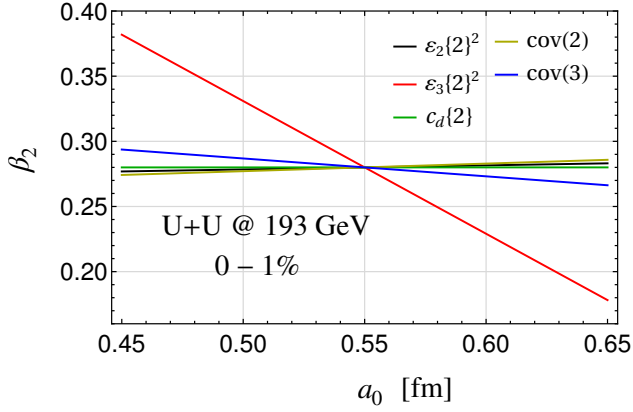
\begin{figure}[t!]
	\includegraphics[scale=.27]{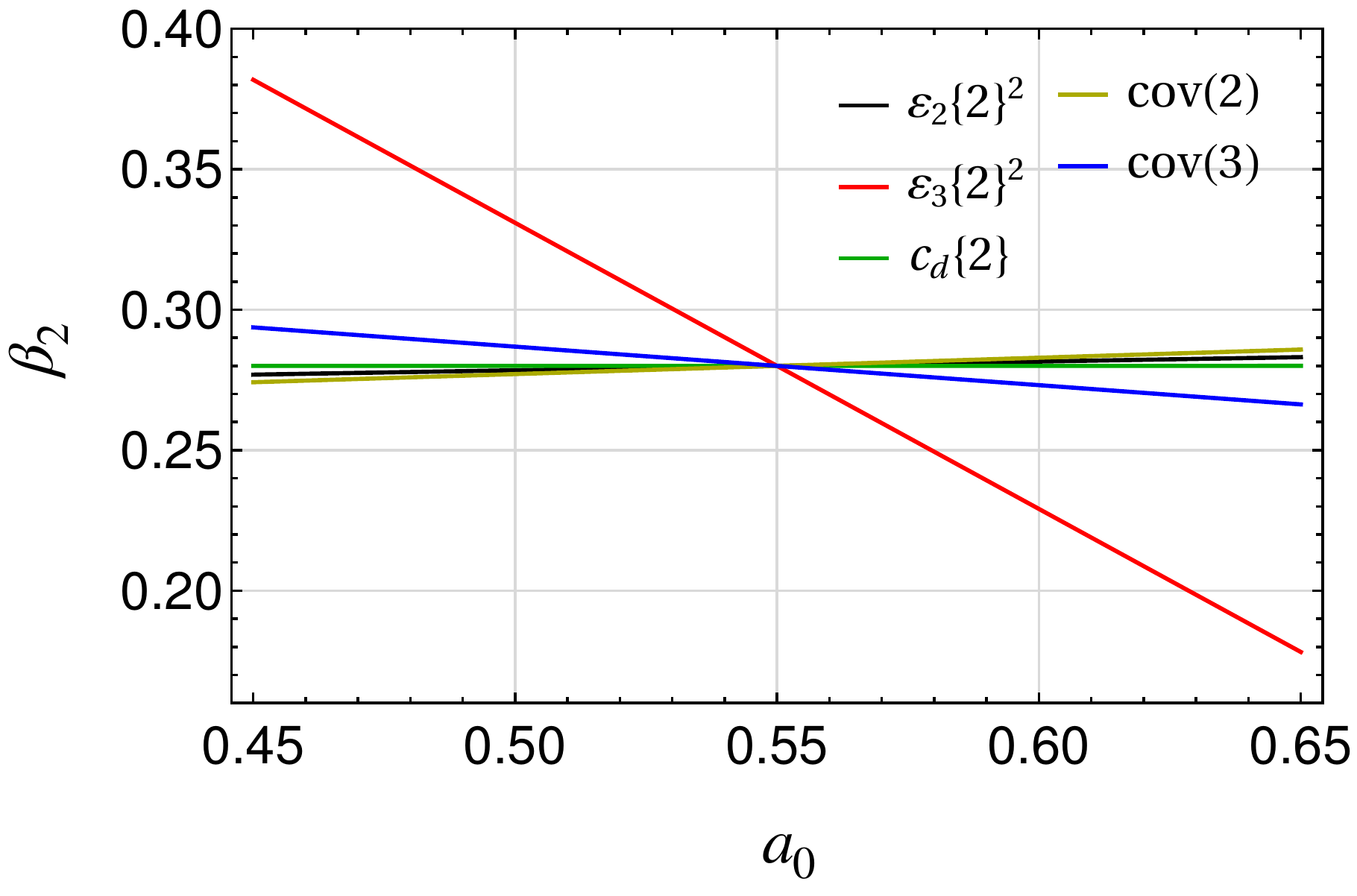}
	\begin{picture}(0,0)
		\put(-180,67){{\fontsize{11}{11}\selectfont \textcolor{black}{U+U @ 193 GeV}}}
		\put(-160,52){{\fontsize{11}{11}\selectfont \textcolor{black}{$0-1\%$}}}
		\put(-90,6){{\fontsize{11}{11}\selectfont \textcolor{black}{[fm]}}}
	\end{picture}	
	\caption{Local deformation--diffuseness degeneracy directions for the five
		initial-state observables at
		$(a_0,\beta_2)=(0.55,0.28)$ for
		$w=d_{\min}=0.5$~fm.
		The curves show the first-order constant-observable directions
		obtained from
		Eq.~\ref{deg}.
		The substantially different slopes demonstrate that the five
		observables probe different combinations of surface diffuseness and
		quadrupole deformation. 
	}	
	\label{fig:local_degenerac}
\end{figure}

The centrality dependence further shows that no single multiparticle
correlation provides a complete characterization of the nuclear surface.
The elliptic two-particle correlation $\varepsilon_2\{2\}$ and the
transverse-size correlation $c_d\{2\}$ are strongly aligned with the
quadrupole response, whereas $\varepsilon_3\{2\}$ and the three-particle
correlations provide complementary sensitivity to the surface profile
and to its interplay with deformation. The relevant question is therefore
not whether a given correlation is sensitive to $a_0$ in isolation, but
whether the combined set of two- and three-particle correlations contains
sufficiently independent response directions to separate $a_0$ from
$\beta_2$. The local degeneracy analysis in the following section
quantifies this complementarity directly in the nuclear-parameter
space.

\begin{figure}[t!]
	\includegraphics[scale=.22]{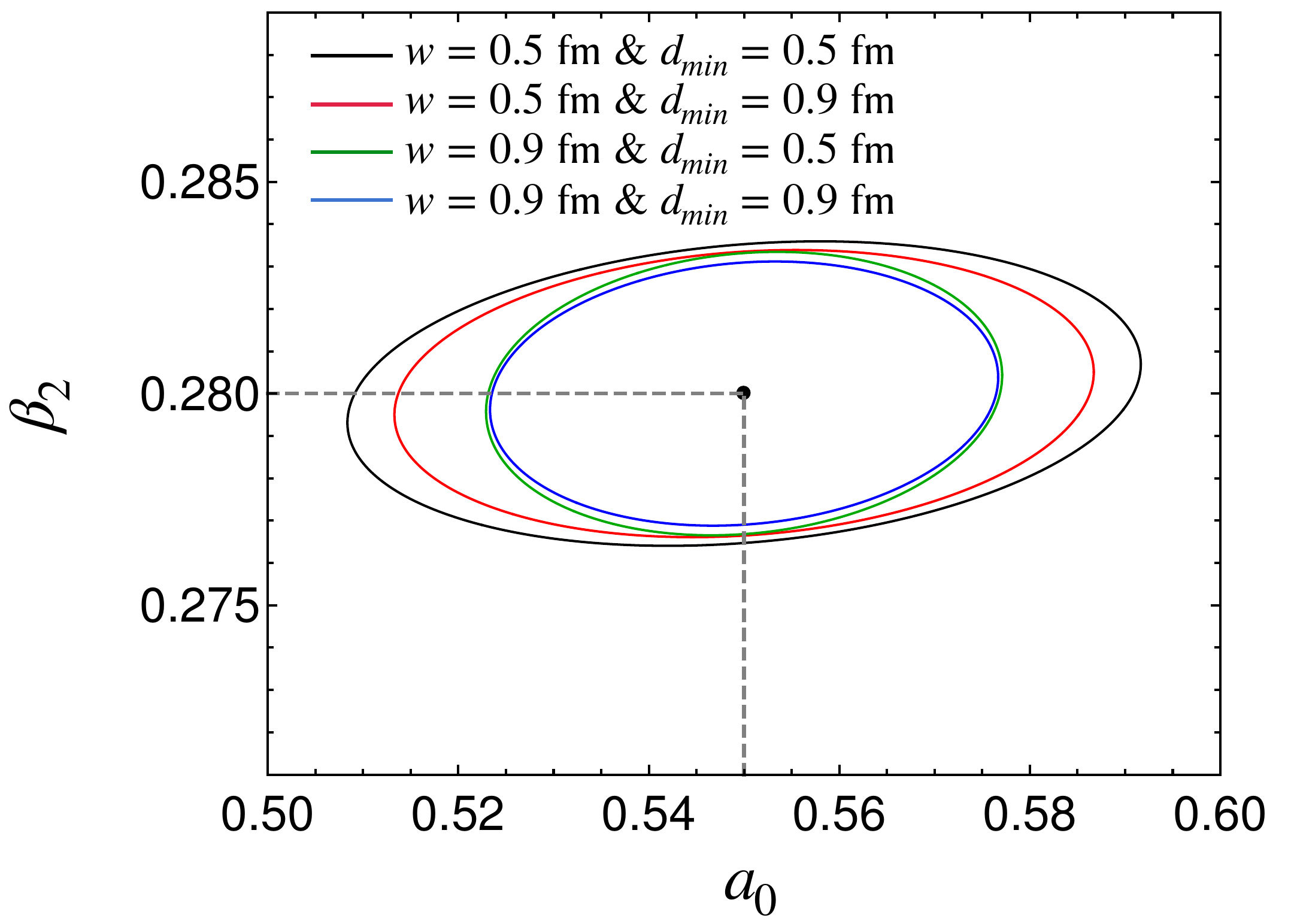}
	\begin{picture}(0,0)
		\put(-98,57){{\fontsize{11}{11}\selectfont \textcolor{black}{U+U @ 193 GeV}}}
		\put(-80,42){{\fontsize{11}{11}\selectfont \textcolor{black}{$0-1\%$}}}
		\put(-85,4){{\fontsize{11}{11}\selectfont \textcolor{black}{[fm]}}}
	\end{picture}	
	\caption{Joint Fisher-information geometry in the $(a_0,\beta_2)$ plane at
		$(a_0,\beta_2)=(0.55,0.28)$.
		The contours correspond to
		$\Delta\chi^2=2.30$, i.e. the joint $68\%$ confidence region for the
		two nuclear-structure parameters, obtained by combining
		$\varepsilon_2\{2\}^2$, $\varepsilon_3\{2\}^2$, $c_d\{2\}$,
		$\mathrm{cov}(2)$, and $\mathrm{cov}(3)$.
		The four curves correspond to different nucleon width and minimum
		inter-nucleon-distance choices.
		Although the size of the constraint varies with the initial-state
		model, all configurations exhibit a strongly elongated parameter
		region, demonstrating that the combined observables retain a
		substantial local degeneracy, predominantly along the $a_0$
		direction.
		The black point indicates the reference parameter set.}	
	\label{fig:fisher_geometry}
\end{figure}
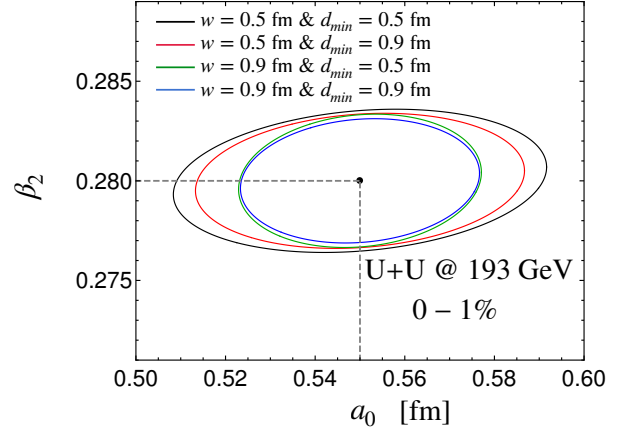
\subsection{Local deformation--diffuseness degeneracy:}
The different responses of the multiparticle correlations to $a_0$ and
$\beta_2$ suggest that their information content can be characterized
locally in the nuclear-parameter space. Although the sensitivity and
information-geometric analyses are performed using the scaled parameter
coordinates introduced in \ref{app:bayesian2D}, the resulting degeneracy directions
can be expressed equivalently in the physical $(a_0,\beta_2)$ coordinates.
For a fixed correlation $X$, variations along a contour of constant
correlation satisfy
\begin{equation}\label{satis}
	dX
	=
	\frac{\partial X}{\partial a_0}\,da_0
	+
	\frac{\partial X}{\partial\beta_2}\,d\beta_2
	=0,
\end{equation}
which defines the local deformation--diffuseness degeneracy direction,
\begin{equation}\label{deg}
	\left.
	\frac{d\beta_2}{da_0}
	\right|_{X}
	=
	-
	\frac{\partial X/\partial a_0}
	{\partial X/\partial\beta_2}.
\end{equation}

\begin{figure*}[t!]
	\includegraphics[scale=.27]{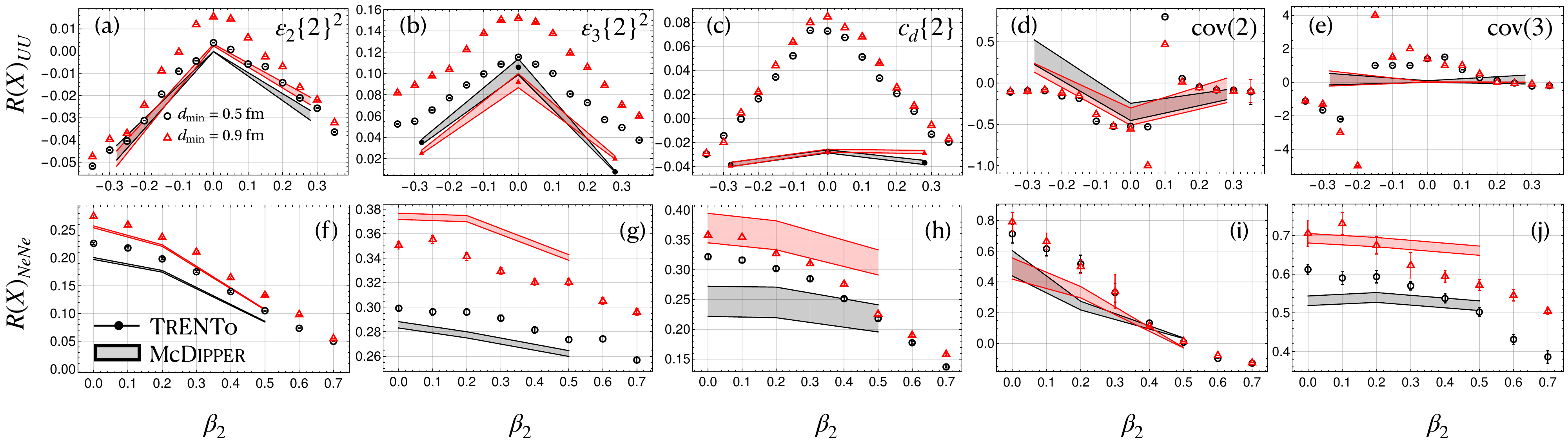}
	\caption{Response of the initial-state observables to quadrupole deformation for
		$^{238}$U+$^{238}$U at 193 GeV (upper row) and $^{20}$Ne+$^{20}$Ne at 5.36 TeV (lower row).
		The columns show $\epsilon_2\{2\}^2$, $\epsilon_3\{2\}^2$,
		$c_d\{2\}$, $\mathrm{cov}(2)$, and $\mathrm{cov}(3)$, respectively.
		For both collisions systems, black circles and red triangles correspond to
		$d_{\min}=0.5$ and $0.9$~fm, respectively, while the corresponding
		colored bands show the results from the \Dipper\; model. The \trento\; ($w=0.5$ fm) results are shown by
		symbols, while the \Dipper\; ($B_G=0.156$ fm$^2$)~\cite{Garcia-Montero:2023gex,Garcia-Montero:2024jev,Garcia-Montero:2025bpn} results are represented by shaded bands.
	}	
	\label{fig:model}
\end{figure*}

Figure~\ref{fig:local_degenerac} shows the resulting local directions
around $(a_0,\beta_2)=(0.55,0.28)$ for $w=d_{\min}=0.5~\mathrm{fm}$. The
substantially different directions obtained from the five multiparticle
correlations demonstrate that the deformation--diffuseness ambiguity is
not a universal property of the nuclear geometry, but depends on the
particular correlation used to probe it.

The two-particle elliptic and transverse-size correlations,
$\varepsilon_2\{2\}$ and $c_d\{2\}$, as well as the three-particle
correlation $\mathrm{cov}(2)$, have nearly horizontal constant-correlation
directions in the $(a_0,\beta_2)$ plane. Their local slopes are
\begin{equation}
	\left\{
	\left.\frac{d\beta_2}{da_0}\right|_{X}
	\right\}
	=
	\{0.031,\,-0.001,\,0.058\},
\end{equation}
respectively. These small values indicate that changes in $a_0$ produce
only a comparatively small shift in the value of $\beta_2$ required to
compensate them, reflecting the dominant sensitivity of these
correlations to the quadrupole deformation in the vicinity of the
reference point. By contrast, $\varepsilon_3\{2\}$ has a substantially
steeper local degeneracy direction,
\begin{equation}
	d\beta_2/da_0
	|_{\varepsilon_3\{2\}}
	=-1.02,
\end{equation}
while $\mathrm{cov}(3)$ gives
\begin{equation}
	d\beta_2/da_0
	|_{\mathrm{cov}(3)}
	=-0.13.
\end{equation}
The opposite signs and different magnitudes of these directions show
that the correlations do not encode the surface diffuseness and
quadrupole deformation in the same combination. In particular, the
triangular correlations provide response directions that differ
substantially from those dominated by elliptic geometry, indicating the
possibility of breaking the deformation--diffuseness degeneracy through
their simultaneous use.

Bayesian parameter estimation has become a standard tool for quantifying
parameter constraints and model uncertainties in relativistic heavy ion
collisions~\cite{Bernhard:2016tnd,Bass:2017jkj,Nijs:2020roc}. We therefore
construct the corresponding Fisher-information matrix
~\cite{LeCam1986,Kass1990,Robert2007},
\begin{equation}\label{fisher}
	F=J^{T}\Sigma^{-1}J,
\end{equation}
where $\Sigma$ contains the full covariance matrix of the five observables,
including the cross-observable covariances evaluated from the common
Monte Carlo event ensemble and the model-discrepancy contribution
discussed in Sec.~\ref{sec:models}. The inverse Fisher matrix,
$C=F^{-1}$, defines the local parameter covariance, while the
constant-information contours satisfy~\cite{Huan:2022}
\begin{equation}
	\Delta\mathbf{x}^{T}F\Delta\mathbf{x}=\Delta\chi^2.
\end{equation}
For the two-dimensional analysis,
$\mathbf{x}=(x_{a_0},x_{\beta_2})$, while the three-dimensional analysis uses
the corresponding $x_{\beta_3}$ coordinate.

Figure~\ref{fig:fisher_geometry} shows the resulting
$\Delta\chi^2=2.30$ contours around
$(a_0,\beta_2)=(0.55,0.28)$ for the four combinations of nucleon width and minimum inter-nucleon separation. Although all five multiparticle correlations are combined, the information contours remain strongly elongated along the $a_0$ direction, demonstrating that the local
degeneracy is not removed simply by increasing the number of correlations. The Fisher condition numbers are $245$, $156$, $98$, and $82$ for $(w,d_{\min})=(0.5,0.5)$, $(0.5,0.9)$, $(0.9,0.5)$, and $(0.9,0.9)~\mathrm{fm}$, respectively. The corresponding weak Fisher eigenvectors are
\begin{equation}
	\mathbf v_{\rm weak}\simeq
	(-1,-0.017),\;
	(-1,-0.012),\;
	(-1,-0.016),\;
	(-1,-0.014),
\end{equation}
respectively, showing that the least-constrained direction is nearly aligned with the surface-diffuseness axis in all four configurations.

This joint information geometry highlights an important distinction
between \emph{complementarity} and \emph{complete identifiability}.
The different multiparticle correlations provide substantially
different local response directions and therefore contain complementary
information, but their combination does not completely eliminate the
deformation--diffuseness ambiguity. The dominant eigenvalue of the
Fisher matrix corresponds to a direction controlled primarily by
$\beta_2$, whereas the much smaller eigenvalue is associated with a
direction dominated by $a_0$. The persistence of this weak-information
direction across the four initial-state configurations demonstrates
that the limitation is not simply a consequence of a particular choice
of nucleon width or short-range nucleon correlations. Instead, it
reflects an intrinsic local degeneracy in how the multiparticle
correlations encode the nuclear surface and quadrupole deformation.

\subsection{Robustness against the nuclear system and initial-state model:}
Having established the deformation--diffuseness degeneracy and the complementary information carried by the multiparticle correlations, we next ask whether the observed response to nuclear deformation is robust against changes in the collision system and initial-state prescription. Figure~\ref{fig:model} compares U+U collisions at $\sqrt{s_{NN}}=193$~GeV with Ne+Ne collisions at $\sqrt{s_{NN}}=5.36$~TeV, using \trento\ with $w=0.5$~fm and $d_{\min}=0.5$ and $0.9$~fm, together with corresponding \Dipper\ calculations. For U+U, the deformation range is approximately $-0.35\lesssim\beta_2\lesssim0.35$, whereas the larger deformation range accessible in Ne+Ne extends to $\beta_2\simeq0.7$\footnote{Unlike heavy nuclei, whose intrinsic structure is commonly described by deformed WS densities, light nuclei can also be described by microscopic \emph{ab initio} calculations that explicitly include nucleon correlations and clustering~\cite{Sun:2026yrr,Lonardoni:2018nob,Frosini:2021ddm,Elhatisari:2017eno,Meissner:2014lgi}. Here we employ a deformed WS parameterization~\cite{ANGELI201369} to isolate the influence of quadrupole deformation on the collective response of the QGP.}. Despite the simultaneous changes in nuclear species and collision energy, the principal $\beta_2$ dependence of the multiparticle correlations remains systematic.

For U+U, changing the minimum inter-nucleon separation produces relatively small modifications of the deformation response, typically at the level of $\mathcal{O}(10^{-2})$ for $\varepsilon_2\{2\}$ and $\varepsilon_3\{2\}$, with somewhat larger effects for the covariance observables. The Ne+Ne comparison provides a more stringent test: the two \trento\ calculations remain qualitatively consistent over the substantially larger deformation range, although their absolute values can differ by $\mathcal{O}(10^{-2})$--$\mathcal{O}(10^{-1})$. The comparison with \Dipper\ likewise shows substantial overlap for several correlations, while deviations of up to several tenths in $R(X)$ can occur for $\mathrm{cov}(2)$ and $\mathrm{cov}(3)$. Thus, the detailed magnitude of the response is observable- and model-dependent, but its systematic deformation dependence is not tied to a particular initial-state construction.

These comparisons demonstrate that the deformation-driven reduction of independent information on $a_0$ is not a peculiarity of U+U collisions, a particular collision energy, or a specific treatment of short-range nucleon correlations. Rather, it reflects the generic mapping of intrinsic nuclear deformation onto multiparticle correlations in relativistic nuclear collisions. At the same time, the quantitative differences between \trento\ and \Dipper\ emphasize that model uncertainties must be incorporated when converting experimentally measured correlations into constraints on the underlying WS parameters.

\subsection{Configuration-dependent sensitivity to quadrupole and octupole deformation:}

\begin{figure}[t!]
	\includegraphics[scale=.34]{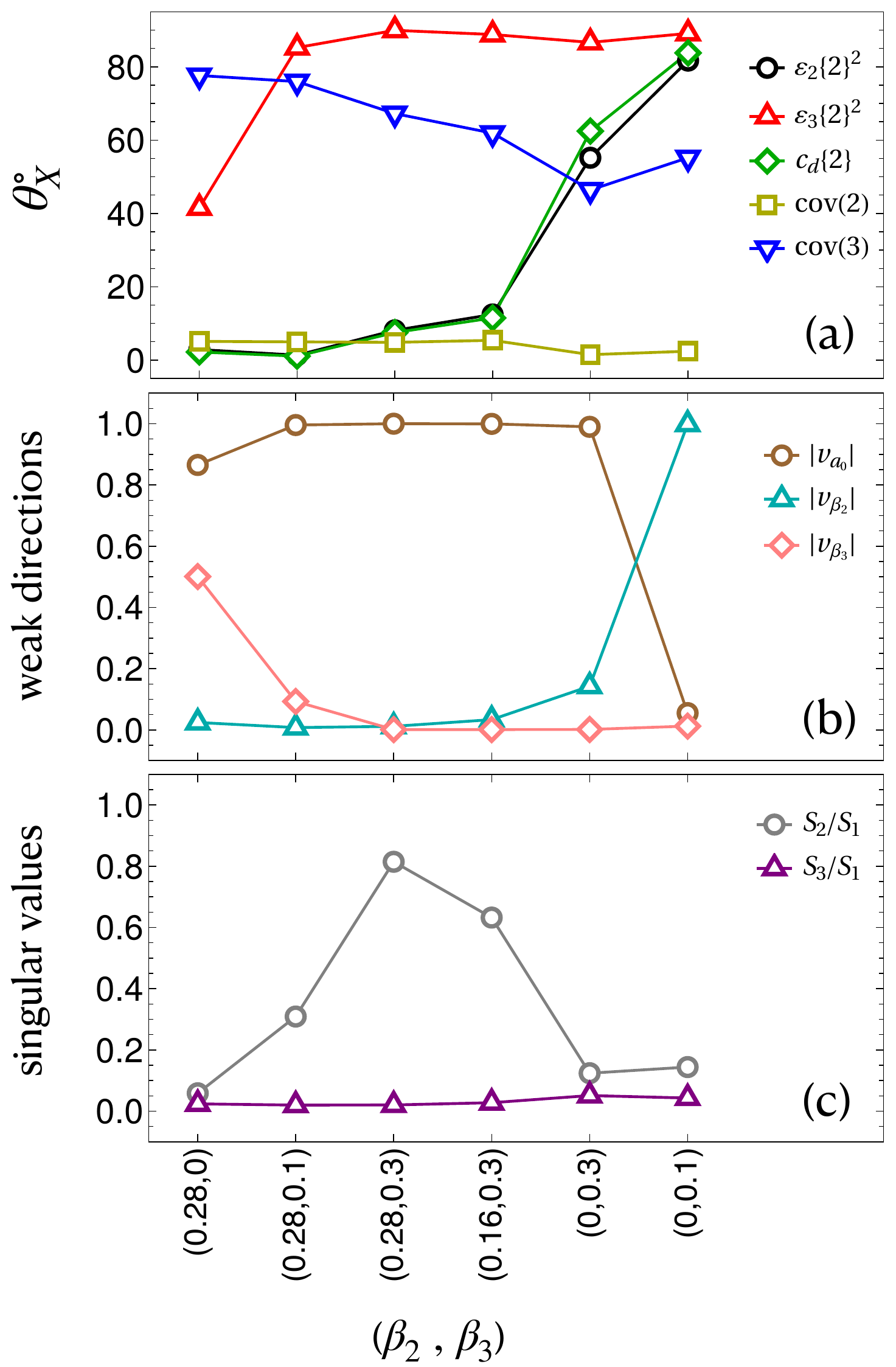}
	\begin{picture}(0,0)
		\put(-190,145){{\fontsize{11}{11}\selectfont \textcolor{black}{U+U @ 193 GeV}}}
		\put(-190,130){{\fontsize{11}{11}\selectfont \textcolor{black}{$0-1\%$}}}
	\end{picture}	
	\caption{
		Configuration dependence of the local sensitivity and information
		geometry in the simultaneous $(\beta_2,\beta_3)$ deformation space
		for $a_0=0.55$ and $w=d_{\min}=0.5~\mathrm{fm}$.
		(a) Local deformation-response angle, Eq.~\ref{theta}, for the five
		initial-state observables.
		(b) Absolute components of the weakest Fisher-information eigenvector,
		showing the evolution of the least-constrained direction in the
		three-dimensional nuclear-parameter space.
		(c) Ratios of the subleading to leading singular values of the
		uncertainty-weighted Jacobian, $S_2/S_1$ and $S_3/S_1$, quantifying
		the hierarchy of locally constrained parameter combinations.
		The six points correspond to the indicated $(\beta_2,\beta_3)$
		configurations.
	}
	\label{fig:beta3}
\end{figure}
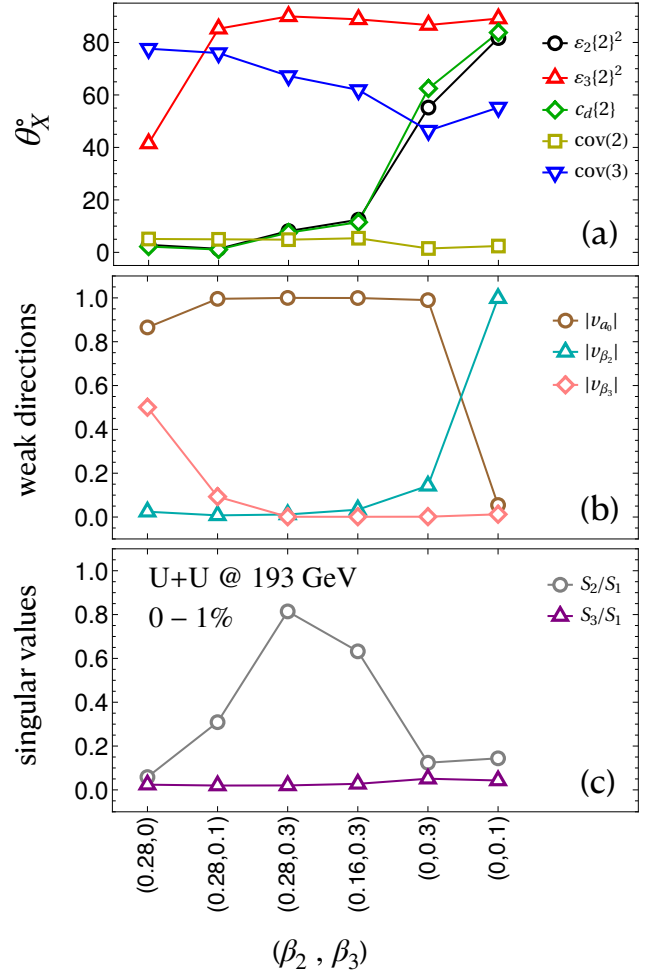

The simultaneous presence of quadrupole and octupole deformation introduces
an additional dimension in the nuclear-parameter space and reveals that the
sensitivity of the multiparticle correlations depends strongly on the
underlying nuclear configuration. Figure~\ref{fig:beta3}(a) shows the local
response angle
\begin{equation}
	\theta_{X}
	=
	\tan^{-1}\left|
	\frac{\partial X/\partial x_{\beta_3}}
	{\partial X/\partial x_{\beta_2}}
	\right|,
	\label{theta}
\end{equation}
which quantifies the relative sensitivity of each correlation to the two
deformation parameters.  For $\varepsilon_3\{2\}$, the response rotates
from a mixed $\beta_2$--$\beta_3$ sensitivity at
$(\beta_2,\beta_3)=(0.28,0)$ toward an almost purely $\beta_3$ response once
finite octupole deformation is introduced, with
$\theta_{\varepsilon_3}$ approaching $90^\circ$. In contrast,
$\varepsilon_2\{2\}$ is initially dominated by $\beta_2$, but its response
acquires an increasingly strong $\beta_3$ component as $\beta_2$ is reduced,
reaching $\theta_{\varepsilon_2}\simeq82^\circ$ at $(0,0.1)$. The other
correlations show smaller but systematic rotations of their response
directions. Thus, the deformation sensitivity of a given multiparticle
correlation cannot be assigned uniquely to $\beta_2$ or $\beta_3$ without
specifying the underlying nuclear configuration.

Figures~\ref{fig:beta3}(b) and \ref{fig:beta3}(c) show how these changes in
local response translate into the information geometry of the full
three-dimensional parameter space. As depicted in Fig.~\ref{fig:beta3}(b),
the weakest Fisher-information direction varies substantially across the
configurations. At $(0.28,0)$ it contains appreciable contributions from
both $a_0$ and $\beta_3$, whereas for $(0.28,0.1)$, $(0.28,0.3)$,
$(0.16,0.3)$, and $(0,0.3)$ it becomes predominantly aligned with $a_0$.
At $(0,0.1)$, however, the weak direction rotates almost entirely toward
$\beta_2$, with $|v_{\beta_2}|\simeq0.998$.

Because the five observables have different magnitudes and uncertainties,
we construct the uncertainty-weighted Jacobian~\cite{Amari2000},
\begin{equation}
	\widetilde{J}
	=
	\Sigma^{-1/2}J.
\end{equation}
The derivatives entering $J$ are evaluated in the same dimensionless
scaled nuclear-parameter coordinates used to construct the response model.
We then perform a singular-value decomposition,
\begin{equation}
	\widetilde{J}
	=
	U\,S\,V^{T},
	\qquad
	S=
	{\rm diag}(S_1,S_2,S_3),
	\qquad
	S_1\geq S_2\geq S_3\geq0.
	\label{eq:svd}
\end{equation}
Here, $S_i$ quantify the strength of the local response to three orthogonal
combinations of the scaled WS parameters, while the corresponding columns
of $V$ define these directions. Thus, $S_1$ identifies the most strongly
constrained local combination, whereas $S_3$ identifies the weakest and
hence least-constrained direction. The scaling convention and its relation
to the physical parameters are described in \ref{app:bayesian}.

As illustrated in Fig.~\ref{fig:beta3}(c), the singular-value hierarchy
exhibits a complementary evolution: $S_2/S_1$ increases from $0.058$ at
$(0.28,0)$ to $0.31$ at $(0.28,0.1)$ and $0.82$ at $(0.28,0.3)$, while
$S_3/S_1$ remains small, between approximately $0.02$ and $0.06$. A finite
octupole deformation therefore substantially strengthens a second
independently constrained parameter combination, without eliminating the
weakest direction.

These results demonstrate that the identifiability of the three WS
parameters is intrinsically configuration dependent. In particular, the
addition of $\beta_3$ does not simply provide a third independent handle
on the nuclear density; instead, it modifies the local orientation of the
response vectors and can redistribute the available information among
$a_0$, $\beta_2$, and $\beta_3$. The simultaneous analysis of multiparticle
correlations is therefore essential for determining which combinations of
nuclear-structure parameters can be constrained in a given region of
parameter space.

\section{Conclusions}\label{sec:conclusions}

We have investigated the information content of multiparticle correlations for constraining the surface diffuseness $a_0$ and intrinsic deformation of nuclei in relativistic heavy ion collisions. Using systematically varied deformed Woods--Saxon configurations, we studied the two-particle correlation observables $\varepsilon_2\{2\}^2$, $\varepsilon_3\{2\}^2$, and $c_d\{2\}$, together with the three-particle flow--transverse-size correlations $\mathrm{cov}(2)$ and $\mathrm{cov}(3)$. The results demonstrate that the extraction of $a_0$ cannot, in general, be separated from the determination of intrinsic deformation. Although all five correlations retain some sensitivity to the surface profile, their responses are strongly configuration and centrality dependent, and in particular the elliptic correlations are dominated by the quadrupole deformation. This produces a local deformation--diffuseness degeneracy in which variations of $a_0$ can be compensated by changes in $\beta_2$. The degeneracy is therefore not a global property of the WS parameter space, but a local feature of the response of the multiparticle correlations.

The different correlations nevertheless contain complementary information. The local Jacobian analysis shows that $\varepsilon_2\{2\}^2$ and $c_d\{2\}$ are predominantly aligned with the $\beta_2$ direction near the reference configuration, whereas $\varepsilon_3\{2\}^2$ and the three-particle covariance observables provide additional sensitivity to the surface diffuseness and to different combinations of the nuclear parameters. The Fisher-information analysis makes this complementarity quantitative: the best-constrained direction is largely associated with quadrupole deformation, while the weakest direction retains a substantial $a_0$ component. Consequently, improved statistical precision alone cannot remove the underlying geometric ambiguity. Our Bayesian analysis reaches the same conclusion after simultaneously accounting for Monte Carlo uncertainties and model discrepancy in the response model, demonstrating that the limitations arise from the structure of the observable response rather than from statistical fluctuations alone.

We have also examined the robustness of these conclusions against changes of the collision system and the initial-state prescription. The comparison of $^{238}$U+$^{238}$U at $\sqrt{s_{NN}}=193$~GeV with $^{20}$Ne+$^{20}$Ne at $\sqrt{s_{NN}}=5.36$~TeV, together with comparisons between \trento\ and \Dipper, shows that the systematic deformation dependence of the multiparticle correlations persists when the nuclear species, collision energy, and treatment of short-range nucleon correlations are changed. The quantitative response of individual correlations can, however, vary appreciably, particularly for the covariance observables. This emphasizes that model uncertainties must be included when translating measured multiparticle correlations into quantitative constraints on nuclear structure.

Finally, the three-dimensional $(a_0,\beta_2,\beta_3)$ analysis reveals that the information geometry itself evolves with the intrinsic deformation. The response directions of the correlations rotate substantially as the $(\beta_2,\beta_3)$ configuration is changed. In particular, $\varepsilon_3\{2\}^2$ becomes predominantly sensitive to $\beta_3$ once finite octupole deformation is introduced, while $\varepsilon_2\{2\}^2$ can acquire significant $\beta_3$ sensitivity as $\beta_2$ is reduced. The weakest Fisher-information direction consequently changes across the deformation space, while the second singular value can increase substantially for finite $\beta_3$, indicating the emergence of an additional independently constrained parameter combination. Thus, there is no universal hierarchy in which a given multiparticle correlation can be assigned uniquely to a single nuclear-structure parameter. Instead, the identifiability of $(a_0,\beta_2,\beta_3)$ is intrinsically configuration dependent. These results establish that relativistic heavy ion collisions can provide valuable information on nuclear many-body structure, but that a reliable extraction of the surface diffuseness requires the simultaneous treatment of intrinsic deformation, complementary multiparticle correlations, and initial-state model uncertainties. Combining such observables with independent low-energy nuclear-structure constraints will therefore be essential for disentangling the surface profile from intrinsic deformation.

\textit{Acknowledgments.—} We thank Björn Schenke, Jiangyong Jia, Gou-Liang Ma, Chunjiang Zhang and You Zhou for useful discussions. H.M is partially supported by the National  Natural Science Foundation of China under Grant No 12347106. 
B.B is supported by research funded by Iran National Science Foundation under Project No. 4040293. O.G-M is supported by European Research Council under project ERC-2018-ADG-835105 YoctoLHC, and by Maria de Maeztu excellent unit grant CEX2023-001318-M. L.Y is supported in part by the National Natural Science Foundation of China through Grants No. 12375133 and No. 12147101. 

\bibliographystyle{elsarticle-num}
\bibliography{biblio.bib}

\begin{thebibliography}{100}
\expandafter\ifx\csname url\endcsname\relax
  \def\url#1{\texttt{#1}}\fi
\expandafter\ifx\csname urlprefix\endcsname\relax\def\urlprefix{URL }\fi
\expandafter\ifx\csname href\endcsname\relax
  \def\href#1#2{#2} \def\path#1{#1}\fi

\bibitem{Bender2003}
M.~Bender, P.-H. Heenen, P.-G. Reinhard, Self-consistent mean-field models for
  nuclear structure, Rev. Mod. Phys. 75 (2003) 121--180.

\bibitem{Schunck2019}
N.~Schunck, Energy density functional methods for atomic nuclei, IOP Publishing
  (2019).

\bibitem{Fricke:1995zz}
G.~Fricke, C.~Bernhardt, K.~Heilig, L.~A. Schaller, L.~Schellenberg, E.~B.
  Shera, C.~W. de~Jager, {Nuclear Ground State Charge Radii from
  Electromagnetic Interactions}, Atom. Data Nucl. Data Tabl. 60 (1995)
  177--285.
\newblock \href {https://doi.org/10.1006/adnd.1995.1007}
  {\path{doi:10.1006/adnd.1995.1007}}.

\bibitem{DeVries1987}
H.~de~Vries, C.~W. de~Jager, C.~de~Vries, Nuclear charge-density-distribution
  parameters from elastic electron scattering, Atomic Data and Nuclear Data
  Tables 36 (1987) 495--536.

\bibitem{Thiel:2019tkm}
M.~Thiel, C.~Sfienti, J.~Piekarewicz, C.~J. Horowitz, M.~Vanderhaeghen,
  {Neutron skins of atomic nuclei: per aspera ad astra}, J. Phys. G 46~(9)
  (2019) 093003.
\newblock \href {http://arxiv.org/abs/1904.12269} {\path{arXiv:1904.12269}},
  \href {https://doi.org/10.1088/1361-6471/ab2c6d}
  {\path{doi:10.1088/1361-6471/ab2c6d}}.

\bibitem{Brown:2000pd}
B.~A. Brown, {Neutron radii in nuclei and the neutron equation of state}, Phys.
  Rev. Lett. 85 (2000) 5296--5299.
\newblock \href {https://doi.org/10.1103/PhysRevLett.85.5296}
  {\path{doi:10.1103/PhysRevLett.85.5296}}.

\bibitem{PREX:2021umo}
D.~Adhikari, et~al., {Accurate Determination of the Neutron Skin Thickness of
  $^{208}$Pb through Parity-Violation in Electron Scattering}, Phys. Rev. Lett.
  126~(17) (2021) 172502.
\newblock \href {http://arxiv.org/abs/2102.10767} {\path{arXiv:2102.10767}},
  \href {https://doi.org/10.1103/PhysRevLett.126.172502}
  {\path{doi:10.1103/PhysRevLett.126.172502}}.

\bibitem{CREX:2022kgg}
D.~Adhikari, et~al., {Precision Determination of the Neutral Weak Form Factor
  of Ca48}, Phys. Rev. Lett. 129~(4) (2022) 042501.
\newblock \href {http://arxiv.org/abs/2205.11593} {\path{arXiv:2205.11593}},
  \href {https://doi.org/10.1103/PhysRevLett.129.042501}
  {\path{doi:10.1103/PhysRevLett.129.042501}}.

\bibitem{Pihan:2025pep}
G.~Pihan, A.~Monnai, B.~Schenke, C.~Shen, {Neutron Skin from Conserved Charge
  Measurements at Collider Experiments}, arxiv preprint (9 2025).
\newblock \href {http://arxiv.org/abs/2509.21644} {\path{arXiv:2509.21644}}.

\bibitem{Zhao:2026zno}
X.-L. Zhao, X.-Y. Xie, Y.~Li, G.-L. Ma, {Constraining the neutron skin of
  $^{208}$Pb with anisotropic flow in Pb+Pb collisions at the LHC}, arXiv
  preprint (3 2026).
\newblock \href {http://arxiv.org/abs/2603.07597} {\path{arXiv:2603.07597}}.

\bibitem{Ryssens:2023fkv}
W.~Ryssens, G.~Giacalone, B.~Schenke, C.~Shen, {Evidence of Hexadecapole
  Deformation in Uranium-238 at the Relativistic Heavy Ion Collider}, Phys.
  Rev. Lett. 130~(21) (2023) 212302.
\newblock \href {http://arxiv.org/abs/2302.13617} {\path{arXiv:2302.13617}},
  \href {https://doi.org/10.1103/PhysRevLett.130.212302}
  {\path{doi:10.1103/PhysRevLett.130.212302}}.

\bibitem{Heinz2013}
U.~Heinz, R.~Snellings, {Collective flow and viscosity in relativistic
  heavy-ion collisions}, Ann. Rev. Nucl. Part. Sci. 63 (2013) 123--151.
\newblock \href {http://arxiv.org/abs/1301.2826} {\path{arXiv:1301.2826}},
  \href {https://doi.org/10.1146/annurev-nucl-102212-170540}
  {\path{doi:10.1146/annurev-nucl-102212-170540}}.

\bibitem{Alver2010}
B.~Alver, G.~Roland, Collision geometry fluctuations and triangular flow in
  heavy-ion collisions, Phys. Rev. C 81 (2010) 054905.

\bibitem{Mehrabpour:2026lhj}
H.~Mehrabpour, Z.~Sheibani, L.~Yan, C.~Zhang, A.~Mirjalili, {Nonlinear
  collective flow reveals the breakdown of quadrupole--hexadecapole scaling in
  heavy ion collisions}, arxiv preprint (7 2026).
\newblock \href {http://arxiv.org/abs/2607.18752} {\path{arXiv:2607.18752}}.

\bibitem{Mehrabpour:2026yuc}
H.~Mehrabpour, G.~Giacalone, M.~W. Luzum, {Triaxial shapes and the angular
  structure of nuclear three-body correlations}, arxiv preprint (4 2026).
\newblock \href {http://arxiv.org/abs/2604.00619} {\path{arXiv:2604.00619}}.

\bibitem{Mehrabpour:2023ign}
H.~Mehrabpour, S.~M.~A. Tabatabaee, {Flow distribution analysis as a probe of
  nuclear deformation}, Phys. Rev. C 108~(3) (2023) 034902.
\newblock \href {http://arxiv.org/abs/2301.07770} {\path{arXiv:2301.07770}},
  \href {https://doi.org/10.1103/PhysRevC.108.034902}
  {\path{doi:10.1103/PhysRevC.108.034902}}.

\bibitem{Mehrabpour:2025rzt}
H.~Mehrabpour, A.~Saha, {Longitudinal flow decorrelations in light ion
  collisions}, Eur. Phys. J. C 85~(11) (2025) 1284.
\newblock \href {http://arxiv.org/abs/2501.14243} {\path{arXiv:2501.14243}},
  \href {https://doi.org/10.1140/epjc/s10052-025-14922-3}
  {\path{doi:10.1140/epjc/s10052-025-14922-3}}.

\bibitem{Parida:2026uld}
T.~Parida, P.~Bo{\.z}ek, {Cumulants of mean transverse momentum and elliptic
  flow in the hydrodynamic model of heavy-ion collisions}, arxiv preprint (5
  2026).
\newblock \href {http://arxiv.org/abs/2605.26737} {\path{arXiv:2605.26737}}.

\bibitem{Giacalone:2026fat}
G.~Giacalone, G.~Nijs, W.~van~der Schee, {Yoctosecond imaging of the ground
  state of $^{129}$Xe at the Large Hadron Collider}, arxiv preprint (6 2026).
\newblock \href {http://arxiv.org/abs/2606.03993} {\path{arXiv:2606.03993}}.

\bibitem{Giacalone:2017dud}
G.~Giacalone, J.~Noronha-Hostler, M.~Luzum, J.-Y. Ollitrault, {Hydrodynamic
  predictions for 5.44 TeV Xe+Xe collisions}, Phys. Rev. C 97~(3) (2018)
  034904.
\newblock \href {http://arxiv.org/abs/1711.08499} {\path{arXiv:1711.08499}},
  \href {https://doi.org/10.1103/PhysRevC.97.034904}
  {\path{doi:10.1103/PhysRevC.97.034904}}.

\bibitem{Rybczynski:2019adt}
M.~Rybczy{\'n}ski, W.~Broniowski, {Glauber Monte Carlo predictions for
  ultrarelativistic collisions with $^{16}$O}, Phys. Rev. C 100~(6) (2019)
  064912.
\newblock \href {http://arxiv.org/abs/1910.09489} {\path{arXiv:1910.09489}},
  \href {https://doi.org/10.1103/PhysRevC.100.064912}
  {\path{doi:10.1103/PhysRevC.100.064912}}.

\bibitem{Summerfield:2021oex}
N.~Summerfield, B.-N. Lu, C.~Plumberg, D.~Lee, J.~Noronha-Hostler, A.~Timmins,
  {$^{16}$O $^{16}$O collisions at energies available at the BNL Relativistic
  Heavy Ion Collider and at the CERN Large Hadron Collider comparing $\alpha$
  clustering versus substructure}, Phys. Rev. C 104~(4) (2021) L041901.
\newblock \href {http://arxiv.org/abs/2103.03345} {\path{arXiv:2103.03345}},
  \href {https://doi.org/10.1103/PhysRevC.104.L041901}
  {\path{doi:10.1103/PhysRevC.104.L041901}}.

\bibitem{Zhang:2021kxj}
C.~Zhang, J.~Jia, {Evidence of Quadrupole and Octupole Deformations in
  Zr96+Zr96 and Ru96+Ru96 Collisions at Ultrarelativistic Energies}, Phys. Rev.
  Lett. 128~(2) (2022) 022301.
\newblock \href {http://arxiv.org/abs/2109.01631} {\path{arXiv:2109.01631}},
  \href {https://doi.org/10.1103/PhysRevLett.128.022301}
  {\path{doi:10.1103/PhysRevLett.128.022301}}.

\bibitem{Xu:2021uar}
H.-j. Xu, W.~Zhao, H.~Li, Y.~Zhou, L.-W. Chen, F.~Wang, {Probing nuclear
  structure with mean transverse momentum in relativistic isobar collisions},
  Phys. Rev. C 108~(1) (2023) L011902.
\newblock \href {http://arxiv.org/abs/2111.14812} {\path{arXiv:2111.14812}},
  \href {https://doi.org/10.1103/PhysRevC.108.L011902}
  {\path{doi:10.1103/PhysRevC.108.L011902}}.

\bibitem{Nijs:2021kvn}
G.~Nijs, W.~van~der Schee, {Inferring nuclear structure from heavy isobar
  collisions using Trajectum}, SciPost Phys. 15~(2) (2023) 041.
\newblock \href {http://arxiv.org/abs/2112.13771} {\path{arXiv:2112.13771}},
  \href {https://doi.org/10.21468/SciPostPhys.15.2.041}
  {\path{doi:10.21468/SciPostPhys.15.2.041}}.

\bibitem{Zhao:2022uhl}
S.~Zhao, H.-j. Xu, Y.-X. Liu, H.~Song, {Probing the nuclear deformation with
  three-particle asymmetric cumulant in RHIC isobar runs}, Phys. Lett. B 839
  (2023) 137838.
\newblock \href {http://arxiv.org/abs/2204.02387} {\path{arXiv:2204.02387}},
  \href {https://doi.org/10.1016/j.physletb.2023.137838}
  {\path{doi:10.1016/j.physletb.2023.137838}}.

\bibitem{Samanta:2023qem}
R.~Samanta, P.~Bo{\.z}ek, {Momentum-dependent flow correlations in deformed
  nuclei at collision energies available at the BNL Relativistic Heavy Ion
  Collider}, Phys. Rev. C 107~(5) (2023) 054916.
\newblock \href {http://arxiv.org/abs/2301.10659} {\path{arXiv:2301.10659}},
  \href {https://doi.org/10.1103/PhysRevC.107.054916}
  {\path{doi:10.1103/PhysRevC.107.054916}}.

\bibitem{Giacalone:2023cet}
G.~Giacalone, G.~Nijs, W.~van~der Schee, {Determination of the Neutron Skin of
  Pb208 from Ultrarelativistic Nuclear Collisions}, Phys. Rev. Lett. 131~(20)
  (2023) 202302.
\newblock \href {http://arxiv.org/abs/2305.00015} {\path{arXiv:2305.00015}},
  \href {https://doi.org/10.1103/PhysRevLett.131.202302}
  {\path{doi:10.1103/PhysRevLett.131.202302}}.

\bibitem{Fortier:2023xxy}
N.~M. Fortier, S.~Jeon, C.~Gale, {Comparisons and predictions for collisions of
  deformed U238 nuclei at sNN=193 GeV}, Phys. Rev. C 111~(1) (2025) 014901.
\newblock \href {http://arxiv.org/abs/2308.09816} {\path{arXiv:2308.09816}},
  \href {https://doi.org/10.1103/PhysRevC.111.014901}
  {\path{doi:10.1103/PhysRevC.111.014901}}.

\bibitem{Xu:2024bdh}
H.-j. Xu, J.~Zhao, F.~Wang, {Hexadecapole Deformation of U238 from Relativistic
  Heavy-Ion Collisions Using a Nonlinear Response Coefficient}, Phys. Rev.
  Lett. 132~(26) (2024) 262301.
\newblock \href {http://arxiv.org/abs/2402.16550} {\path{arXiv:2402.16550}},
  \href {https://doi.org/10.1103/PhysRevLett.132.262301}
  {\path{doi:10.1103/PhysRevLett.132.262301}}.

\bibitem{Zhang:2024vkh}
C.~Zhang, J.~Chen, G.~Giacalone, S.~Huang, J.~Jia, Y.-G. Ma, {Ab-initio
  nucleon-nucleon correlations and their impact on high energy 16O+16O
  collisions}, Phys. Lett. B 862 (2025) 139322.
\newblock \href {http://arxiv.org/abs/2404.08385} {\path{arXiv:2404.08385}},
  \href {https://doi.org/10.1016/j.physletb.2025.139322}
  {\path{doi:10.1016/j.physletb.2025.139322}}.

\bibitem{Zhao:2024feh}
X.-L. Zhao, Z.-W. Lin, Y.~Zhou, C.~Zhang, G.-L. Ma, {Nuclear cluster structure
  effect in 16O+16O collisions at the top RHIC energy}, Phys. Lett. B 874
  (2026) 140254.
\newblock \href {http://arxiv.org/abs/2404.09780} {\path{arXiv:2404.09780}},
  \href {https://doi.org/10.1016/j.physletb.2026.140254}
  {\path{doi:10.1016/j.physletb.2026.140254}}.

\bibitem{Fortier:2024yxs}
N.~M. Fortier, S.~Jeon, C.~Gale, {Heavy-ion collisions as probes of nuclear
  structure}, Phys. Rev. C 111~(1) (2025) L011901.
\newblock \href {http://arxiv.org/abs/2405.17526} {\path{arXiv:2405.17526}},
  \href {https://doi.org/10.1103/PhysRevC.111.L011901}
  {\path{doi:10.1103/PhysRevC.111.L011901}}.

\bibitem{Giacalone:2024ixe}
G.~Giacalone, et~al., {Anisotropic Flow in Fixed-Target Pb208+Ne20 Collisions
  as a Probe of Quark-Gluon Plasma}, Phys. Rev. Lett. 134~(8) (2025) 082301.
\newblock \href {http://arxiv.org/abs/2405.20210} {\path{arXiv:2405.20210}},
  \href {https://doi.org/10.1103/PhysRevLett.134.082301}
  {\path{doi:10.1103/PhysRevLett.134.082301}}.

\bibitem{Mantysaari:2024uwn}
H.~M{\"a}ntysaari, B.~Schenke, C.~Shen, W.~Zhao, {Probing nuclear structure of
  heavy ions at energies available at the CERN Large Hadron Collider}, Phys.
  Rev. C 110~(5) (2024) 054913.
\newblock \href {http://arxiv.org/abs/2409.19064} {\path{arXiv:2409.19064}},
  \href {https://doi.org/10.1103/PhysRevC.110.054913}
  {\path{doi:10.1103/PhysRevC.110.054913}}.

\bibitem{Lu:2025cni}
Z.~Lu, M.~Zhao, E.~G.~D. Nielsen, X.~Li, Y.~Zhou, {Imprint of light nuclei
  structure in relativistic nuclear collisions}, Phys. Lett. B 868 (2025)
  139698.
\newblock \href {http://arxiv.org/abs/2501.14852} {\path{arXiv:2501.14852}},
  \href {https://doi.org/10.1016/j.physletb.2025.139698}
  {\path{doi:10.1016/j.physletb.2025.139698}}.

\bibitem{Li:2025vdp}
Y.~Li, X.~Zhang, G.~Giacalone, J.~Yao, {Benchmarking Nuclear Matrix Elements of
  0{\ensuremath{\nu}}{\ensuremath{\beta}}{\ensuremath{\beta}} Decay with
  High-Energy Nuclear Collisions}, Phys. Rev. Lett. 135~(2) (2025) 022301.
\newblock \href {http://arxiv.org/abs/2502.08027} {\path{arXiv:2502.08027}},
  \href {https://doi.org/10.1103/zymp-tyjj} {\path{doi:10.1103/zymp-tyjj}}.

\bibitem{Liu:2025zsi}
L.-M. Liu, H.-C. Wang, S.-J. Li, C.~Zhang, J.~Xu, Z.-Z. Ren, J.~Jia, X.-G.
  Huang, {Directly probing existence of {\ensuremath{\alpha}}-cluster structure
  in Ne20 by relativistic heavy-ion collisions}, Phys. Rev. C 111~(2) (2025)
  L021901.
\newblock \href {http://arxiv.org/abs/2502.08057} {\path{arXiv:2502.08057}},
  \href {https://doi.org/10.1103/PhysRevC.111.L021901}
  {\path{doi:10.1103/PhysRevC.111.L021901}}.

\bibitem{Li:2025hae}
P.~Li, B.~Zhou, G.-L. Ma, {Identifying {\ensuremath{\alpha}}-Cluster
  Configurations in Ne20 via Ultracentral Ne+Ne Collisions}, Phys. Rev. Lett.
  136~(8) (2026) 082302.
\newblock \href {http://arxiv.org/abs/2504.04688} {\path{arXiv:2504.04688}},
  \href {https://doi.org/10.1103/tffz-8q1m} {\path{doi:10.1103/tffz-8q1m}}.

\bibitem{Zhang:2025hvi}
C.~Zhang, J.~Jia, J.~Chen, C.~Shen, L.~Liu, {Probing the octupole deformation
  of $^{238}$U in high-energy nuclear collisions}, arxiv preprint (4 2025).
\newblock \href {http://arxiv.org/abs/2504.15245} {\path{arXiv:2504.15245}}.

\bibitem{TabatabaeeMehr:2024lgu}
S.~M.~A. Tabatabaee~Mehr, S.~F. Taghavi, {Revealing initial-state properties
  through ultracentral symmetric heavy-ion collisions}, Phys. Rev. C 110~(5)
  (2024) 054901.
\newblock \href {http://arxiv.org/abs/2406.13863} {\path{arXiv:2406.13863}},
  \href {https://doi.org/10.1103/PhysRevC.110.054901}
  {\path{doi:10.1103/PhysRevC.110.054901}}.

\bibitem{Taghavi:2025ddm}
S.~F. Taghavi, S.~M.~A. Tabatabaee~Mehr, {Opacity estimation of OO collision
  from CoMBolt-ITA hybrid}, Phys. Lett. B 878 (2026) 140505.
\newblock \href {http://arxiv.org/abs/2512.05009} {\path{arXiv:2512.05009}},
  \href {https://doi.org/10.1016/j.physletb.2026.140505}
  {\path{doi:10.1016/j.physletb.2026.140505}}.

\bibitem{Li:2026igf}
Y.~Li, H.-j. Xu, D.~Zhang, G.-L. Ma, {Investigation of the shape of uranium in
  relativistic $^{238}$U+$^{238}$U collisions with nuclear densities from
  covariant density functional theory}, arxiv preprint (2 2026).
\newblock \href {http://arxiv.org/abs/2602.02336} {\path{arXiv:2602.02336}}.

\bibitem{Bofos:2026nmg}
S.~Bofos, Y.~Li, C.~Ding, B.~Bally, T.~Duguet, M.~Frosini, J.~Yao, {Quantum
  effects in the quadrupole rotor picture of ultra-relativistic ion-ion
  collisions}, arxiv preprint (5 2026).
\newblock \href {http://arxiv.org/abs/2605.28813} {\path{arXiv:2605.28813}}.

\bibitem{Mehrabpour:2025ogw}
H.~Mehrabpour, {Imprint of {\ensuremath{\alpha}} clustering on ab initio
  correlations in relativistic light ion collisions}, Phys. Rev. C 113~(3)
  (2026) 034909.
\newblock \href {http://arxiv.org/abs/2506.12673} {\path{arXiv:2506.12673}},
  \href {https://doi.org/10.1103/cgcy-j883} {\path{doi:10.1103/cgcy-j883}}.

\bibitem{Liu:2025uks}
Q.~Liu, H.~Mehrabpour, B.-N. Lu, {Impacts of isolated nucleon-nucleon
  correlations in relativistic $^{16}$O+$^{16}$O collisions}, arxiv preprint (8
  2025).
\newblock \href {http://arxiv.org/abs/2509.00315} {\path{arXiv:2509.00315}}.

\bibitem{Giacalone:2023hwk}
G.~Giacalone, {Many-body correlations for nuclear physics across scales: from
  nuclei to quark-gluon plasmas to hadron distributions}, Eur. Phys. J. A
  59~(12) (2023) 297.
\newblock \href {http://arxiv.org/abs/2305.19843} {\path{arXiv:2305.19843}},
  \href {https://doi.org/10.1140/epja/s10050-023-01200-7}
  {\path{doi:10.1140/epja/s10050-023-01200-7}}.

\bibitem{Blaizot:2025bfu}
J.-P. Blaizot, G.~Giacalone, A.~Lovato, {Nuclear collectivity and the harmonic
  spectrum of two-body correlations}, arXiv preprint (12 2025).
\newblock \href {http://arxiv.org/abs/2512.18926} {\path{arXiv:2512.18926}}.

\bibitem{Duguet:2025qxi}
T.~Duguet, G.~Giacalone, V.~Som{\`a}, Y.~Zhou, {Topical issue on the
  intersection of low-energy nuclear structure and high-energy nuclear
  collisions}, Eur. Phys. J. A 61 (2025) 237.
\newblock \href {http://arxiv.org/abs/2512.05874} {\path{arXiv:2512.05874}},
  \href {https://doi.org/10.1140/epja/s10050-025-01715-1}
  {\path{doi:10.1140/epja/s10050-025-01715-1}}.

\bibitem{Blaizot:2025scr}
J.-P. Blaizot, G.~Giacalone, {Angular structure of many-body correlations in
  atomic nuclei}, Eur. Phys. J. A 61~(9) (2025) 220.
\newblock \href {http://arxiv.org/abs/2504.15421} {\path{arXiv:2504.15421}},
  \href {https://doi.org/10.1140/epja/s10050-025-01679-2}
  {\path{doi:10.1140/epja/s10050-025-01679-2}}.

\bibitem{Blaizot:2026yvx}
J.~P. Blaizot, {From Wounded Nucleons to Nuclear Structure}, Acta Phys. Polon.
  B 57~(6) (2026) 6--A8.
\newblock \href {https://doi.org/10.5506/APhysPolB.57.6-A8}
  {\path{doi:10.5506/APhysPolB.57.6-A8}}.

\bibitem{Jia2022b}
J.~Jia, G.~Giacalone, C.~Zhang, {Separating the Impact of Nuclear Skin and
  Nuclear Deformation in High-Energy Isobar Collisions}, Phys. Rev. Lett.
  131~(2) (2023) 022301.
\newblock \href {http://arxiv.org/abs/2206.10449} {\path{arXiv:2206.10449}},
  \href {https://doi.org/10.1103/PhysRevLett.131.022301}
  {\path{doi:10.1103/PhysRevLett.131.022301}}.

\bibitem{Giacalone2023}
G.~Giacalone, G.~Nijs, W.~van~der Schee, {Determination of the Neutron Skin of
  Pb208 from Ultrarelativistic Nuclear Collisions}, Phys. Rev. Lett. 131~(20)
  (2023) 202302.
\newblock \href {http://arxiv.org/abs/2305.00015} {\path{arXiv:2305.00015}},
  \href {https://doi.org/10.1103/PhysRevLett.131.202302}
  {\path{doi:10.1103/PhysRevLett.131.202302}}.

\bibitem{Vitsos:2025jzt}
A.~Vitsos, L.~M.~M. Soranzo, E.~G.~D. Nielsen, Y.~Zhou, {Characterizing the
  neutron skin of $^{48}$Ca through collective flow at the CERN large hadron
  collider}, Eur. Phys. J. C 86~(5) (2026) 536.
\newblock \href {http://arxiv.org/abs/2512.00114} {\path{arXiv:2512.00114}},
  \href {https://doi.org/10.1140/epjc/s10052-026-15751-8}
  {\path{doi:10.1140/epjc/s10052-026-15751-8}}.

\bibitem{Sun:2026yrr}
X.~Sun, J.~Dobaczewski, W.~Nazarewicz, H.~Wibowo, {Multipole tomography of
  atomic nuclei with symmetry-conserved theories}, arXiv preprint (5 2026).
\newblock \href {http://arxiv.org/abs/2605.26088} {\path{arXiv:2605.26088}}.

\bibitem{Li:2026yzw}
H.~Li, L.-M. Liu, J.~Chen, Y.-G. Ma, C.~Zhang, {Probing the Neutron-Skin
  Thickness Through J/{\ensuremath{\psi}} Photoproduction in Ultra-Peripheral
  Collisions}, Chin. Phys. Lett. 43~(5) (2026) 050101.
\newblock \href {http://arxiv.org/abs/2604.19650} {\path{arXiv:2604.19650}},
  \href {https://doi.org/10.1088/0256-307X/43/5/050101}
  {\path{doi:10.1088/0256-307X/43/5/050101}}.

\bibitem{Giacalone:2025vxa}
G.~Giacalone, et~al., {Nuclear Physics Confronts Relativistic Collisions Of
  Isobars}, arXiv preprint (7 2025).
\newblock \href {http://arxiv.org/abs/2507.01454} {\path{arXiv:2507.01454}}.

\bibitem{STAR:2025elk}
{Imaging nuclear shape through anisotropic and radial flow in high-energy
  heavy-ion collisions}, Rept. Prog. Phys. 88~(10) (2025) 108601.
\newblock \href {http://arxiv.org/abs/2506.17785} {\path{arXiv:2506.17785}},
  \href {https://doi.org/10.1088/1361-6633/ae0fc3}
  {\path{doi:10.1088/1361-6633/ae0fc3}}.

\bibitem{Giacalone:2024luz}
G.~Giacalone, et~al., {Exploiting Ne20 Isotopes for Precision Characterizations
  of Collectivity in Small Systems}, Phys. Rev. Lett. 135~(1) (2025) 012302.
\newblock \href {http://arxiv.org/abs/2402.05995} {\path{arXiv:2402.05995}},
  \href {https://doi.org/10.1103/k8rb-jgvq} {\path{doi:10.1103/k8rb-jgvq}}.

\bibitem{Raman2001}
S.~Raman, C.~W.~G. Nestor, Jr, P.~Tikkanen, {Transition probability from the
  ground to the first-excited 2+ state of even-even nuclides}, Atom. Data Nucl.
  Data Tabl. 78 (2001) 1--128.
\newblock \href {https://doi.org/10.1006/adnd.2001.0858}
  {\path{doi:10.1006/adnd.2001.0858}}.

\bibitem{Moreland:2014oya}
J.~S. Moreland, J.~E. Bernhard, S.~A. Bass, {Alternative ansatz to wounded
  nucleon and binary collision scaling in high-energy nuclear collisions},
  Phys. Rev. C 92~(1) (2015) 011901.
\newblock \href {http://arxiv.org/abs/1412.4708} {\path{arXiv:1412.4708}},
  \href {https://doi.org/10.1103/PhysRevC.92.011901}
  {\path{doi:10.1103/PhysRevC.92.011901}}.

\bibitem{Garcia-Montero:2023gex}
O.~Garcia-Montero, H.~Elfner, S.~Schlichting, {McDIPPER: A novel
  saturation-based 3+1D initial-state model for heavy ion collisions}, Phys.
  Rev. C 109~(4) (2024) 044916.
\newblock \href {http://arxiv.org/abs/2308.11713} {\path{arXiv:2308.11713}},
  \href {https://doi.org/10.1103/PhysRevC.109.044916}
  {\path{doi:10.1103/PhysRevC.109.044916}}.

\bibitem{Garcia-Montero:2024jev}
O.~Garcia-Montero, S.~Schlichting, {Baryon stopping and charge deposition in
  heavy-ion collisions due to gluon saturation}, Phys. Rev. C 111~(2) (2025)
  024912.
\newblock \href {http://arxiv.org/abs/2409.06788} {\path{arXiv:2409.06788}},
  \href {https://doi.org/10.1103/PhysRevC.111.024912}
  {\path{doi:10.1103/PhysRevC.111.024912}}.

\bibitem{Garcia-Montero:2025bpn}
O.~Garcia-Montero, S.~Schlichting, J.~Zhu, {Effects of subnucleonic
  fluctuations on the longitudinal structure of heavy-ion collisions}, Phys.
  Rev. D 111~(7) (2025) 076029.
\newblock \href {http://arxiv.org/abs/2501.14872} {\path{arXiv:2501.14872}},
  \href {https://doi.org/10.1103/PhysRevD.111.076029}
  {\path{doi:10.1103/PhysRevD.111.076029}}.

\bibitem{Garcia-Montero:2025hys}
O.~Garcia-Montero, S.~Schlichting, {Effective theories for nuclei at high
  energies}, Eur. Phys. J. A 61~(3) (2025) 54.
\newblock \href {http://arxiv.org/abs/2502.09721} {\path{arXiv:2502.09721}},
  \href {https://doi.org/10.1140/epja/s10050-025-01523-7}
  {\path{doi:10.1140/epja/s10050-025-01523-7}}.

\bibitem{Rose:2014fba}
J.-B. Rose, J.-F. Paquet, G.~S. Denicol, M.~Luzum, B.~Schenke, S.~Jeon,
  C.~Gale, {Extracting the bulk viscosity of the quark{\textendash}gluon
  plasma}, Nucl. Phys. A 931 (2014) 926--930.
\newblock \href {http://arxiv.org/abs/1408.0024} {\path{arXiv:1408.0024}},
  \href {https://doi.org/10.1016/j.nuclphysa.2014.09.044}
  {\path{doi:10.1016/j.nuclphysa.2014.09.044}}.

\bibitem{Ryu:2015vwa}
S.~Ryu, J.~F. Paquet, C.~Shen, G.~S. Denicol, B.~Schenke, S.~Jeon, C.~Gale,
  {Importance of the Bulk Viscosity of QCD in Ultrarelativistic Heavy-Ion
  Collisions}, Phys. Rev. Lett. 115~(13) (2015) 132301.
\newblock \href {http://arxiv.org/abs/1502.01675} {\path{arXiv:1502.01675}},
  \href {https://doi.org/10.1103/PhysRevLett.115.132301}
  {\path{doi:10.1103/PhysRevLett.115.132301}}.

\bibitem{Ryu:2017qzn}
S.~Ryu, J.-F. Paquet, C.~Shen, G.~Denicol, B.~Schenke, S.~Jeon, C.~Gale,
  {Effects of bulk viscosity and hadronic rescattering in heavy ion collisions
  at energies available at the BNL Relativistic Heavy Ion Collider and at the
  CERN Large Hadron Collider}, Phys. Rev. C 97~(3) (2018) 034910.
\newblock \href {http://arxiv.org/abs/1704.04216} {\path{arXiv:1704.04216}},
  \href {https://doi.org/10.1103/PhysRevC.97.034910}
  {\path{doi:10.1103/PhysRevC.97.034910}}.

\bibitem{Miller:2007ri}
M.~L. Miller, K.~Reygers, S.~J. Sanders, P.~Steinberg, {Glauber modeling in
  high energy nuclear collisions}, Ann. Rev. Nucl. Part. Sci. 57 (2007)
  205--243.
\newblock \href {http://arxiv.org/abs/nucl-ex/0701025}
  {\path{arXiv:nucl-ex/0701025}}, \href
  {https://doi.org/10.1146/annurev.nucl.57.090506.123020}
  {\path{doi:10.1146/annurev.nucl.57.090506.123020}}.

\bibitem{PhysRev.95.577}
R.~D. Woods, D.~S. Saxon,
  \href{https://link.aps.org/doi/10.1103/PhysRev.95.577}{Diffuse surface
  optical model for nucleon-nuclei scattering}, Phys. Rev. 95 (1954) 577--578.
\newblock \href {https://doi.org/10.1103/PhysRev.95.577}
  {\path{doi:10.1103/PhysRev.95.577}}.
\newline\urlprefix\url{https://link.aps.org/doi/10.1103/PhysRev.95.577}

\bibitem{BohrMottelson1975}
A.~Bohr, B.~R. Mottelson, Nuclear Structure, Volume II: Nuclear Deformations,
  W. A. Benjamin, Reading, Massachusetts, 1975.

\bibitem{RingSchuck1980}
P.~Ring, P.~Schuck, The Nuclear Many-Body Problem, Springer, New York, 1980.
\newblock \href {https://doi.org/10.1007/978-3-642-61852-9}
  {\path{doi:10.1007/978-3-642-61852-9}}.

\bibitem{Li2024Rn224}
H.~Li, J.~Xiang, J.~Meng, et~al., Static or dynamic pear shapes in radioactive
  nucleus $^{224}$rn?, Nuclear Science and Techniques 35 (2024) 184.
\newblock \href {https://doi.org/10.1007/s41365-024-01578-z}
  {\path{doi:10.1007/s41365-024-01578-z}}.

\bibitem{Robledo2025History}
L.~M. Robledo, History of the concept of nuclear shape, European Physical
  Journal A 61 (2025) 90.
\newblock \href {https://doi.org/10.1140/epja/s10050-025-01545-1}
  {\path{doi:10.1140/epja/s10050-025-01545-1}}.

\bibitem{STAR:2024wgy}
M.~I. Abdulhamid, et~al., {Imaging shapes of atomic nuclei in high-energy
  nuclear collisions}, Nature 635~(8037) (2024) 67--72.
\newblock \href {http://arxiv.org/abs/2401.06625} {\path{arXiv:2401.06625}},
  \href {https://doi.org/10.1038/s41586-024-08097-2}
  {\path{doi:10.1038/s41586-024-08097-2}}.

\bibitem{ANGELI201369}
I.~Angeli, K.~Marinova,
  \href{https://www.sciencedirect.com/science/article/pii/S0092640X12000265}{Table
  of experimental nuclear ground state charge radii: An update}, Atomic Data
  and Nuclear Data Tables 99~(1) (2013) 69--95.
\newblock \href {https://doi.org/https://doi.org/10.1016/j.adt.2011.12.006}
  {\path{doi:https://doi.org/10.1016/j.adt.2011.12.006}}.
\newline\urlprefix\url{https://www.sciencedirect.com/science/article/pii/S0092640X12000265}

\bibitem{Zhang:2025zrm}
C.~Zhang, J.~Jia, J.~Chen, C.~Shen, L.~Liu, {Imprints of octupole collectivity
  in uranium-238 on relativistic heavy-ion flow observables}, Phys. Rev. Res.
  8~(3) (2026) 033018.
\newblock \href {http://arxiv.org/abs/2504.15245} {\path{arXiv:2504.15245}},
  \href {https://doi.org/10.1103/3n5q-m2kf} {\path{doi:10.1103/3n5q-m2kf}}.

\bibitem{Jia:2021qyu}
J.~Jia, {Probing triaxial deformation of atomic nuclei in high-energy heavy ion
  collisions}, Phys. Rev. C 105~(4) (2022) 044905.
\newblock \href {http://arxiv.org/abs/2109.00604} {\path{arXiv:2109.00604}},
  \href {https://doi.org/10.1103/PhysRevC.105.044905}
  {\path{doi:10.1103/PhysRevC.105.044905}}.

\bibitem{Niemi2013}
H.~Niemi, G.~S. Denicol, H.~Holopainen, P.~Huovinen, {Event-by-event
  distributions of azimuthal asymmetries in ultrarelativistic heavy-ion
  collisions}, Phys. Rev. C 87~(5) (2013) 054901.
\newblock \href {http://arxiv.org/abs/1212.1008} {\path{arXiv:1212.1008}},
  \href {https://doi.org/10.1103/PhysRevC.87.054901}
  {\path{doi:10.1103/PhysRevC.87.054901}}.

\bibitem{Teaney:2013dta}
D.~Teaney, L.~Yan, {Event-plane correlations and hydrodynamic simulations of
  heavy ion collisions}, Phys. Rev. C 90~(2) (2014) 024902.
\newblock \href {http://arxiv.org/abs/1312.3689} {\path{arXiv:1312.3689}},
  \href {https://doi.org/10.1103/PhysRevC.90.024902}
  {\path{doi:10.1103/PhysRevC.90.024902}}.

\bibitem{Niemi:2012aj}
H.~Niemi, G.~S. Denicol, H.~Holopainen, P.~Huovinen, {Event-by-event
  distributions of azimuthal asymmetries in ultrarelativistic heavy-ion
  collisions}, Phys. Rev. C 87~(5) (2013) 054901.
\newblock \href {http://arxiv.org/abs/1212.1008} {\path{arXiv:1212.1008}},
  \href {https://doi.org/10.1103/PhysRevC.87.054901}
  {\path{doi:10.1103/PhysRevC.87.054901}}.

\bibitem{Giacalone:2020dln}
G.~Giacalone, F.~G. Gardim, J.~Noronha-Hostler, J.-Y. Ollitrault, {Correlation
  between mean transverse momentum and anisotropic flow in heavy-ion
  collisions}, Phys. Rev. C 103~(2) (2021) 024909.
\newblock \href {http://arxiv.org/abs/2004.01765} {\path{arXiv:2004.01765}},
  \href {https://doi.org/10.1103/PhysRevC.103.024909}
  {\path{doi:10.1103/PhysRevC.103.024909}}.

\bibitem{Jia:2021tzt}
J.~Jia, {Shape of atomic nuclei in heavy ion collisions}, Phys. Rev. C 105~(1)
  (2022) 014905.
\newblock \href {http://arxiv.org/abs/2106.08768} {\path{arXiv:2106.08768}},
  \href {https://doi.org/10.1103/PhysRevC.105.014905}
  {\path{doi:10.1103/PhysRevC.105.014905}}.

\bibitem{Bozek:2016yoj}
P.~Bozek, {Transverse-momentum{\textendash}flow correlations in relativistic
  heavy-ion collisions}, Phys. Rev. C 93~(4) (2016) 044908.
\newblock \href {http://arxiv.org/abs/1601.04513} {\path{arXiv:1601.04513}},
  \href {https://doi.org/10.1103/PhysRevC.93.044908}
  {\path{doi:10.1103/PhysRevC.93.044908}}.

\bibitem{Bozek:2020drh}
P.~Bozek, H.~Mehrabpour, {Correlation coefficient between harmonic flow and
  transverse momentum in heavy-ion collisions}, Phys. Rev. C 101~(6) (2020)
  064902.
\newblock \href {http://arxiv.org/abs/2002.08832} {\path{arXiv:2002.08832}},
  \href {https://doi.org/10.1103/PhysRevC.101.064902}
  {\path{doi:10.1103/PhysRevC.101.064902}}.

\bibitem{Schenke:2020uqq}
B.~Schenke, C.~Shen, D.~Teaney, {Transverse momentum fluctuations and their
  correlation with elliptic flow in nuclear collision}, Phys. Rev. C 102~(3)
  (2020) 034905.
\newblock \href {http://arxiv.org/abs/2004.00690} {\path{arXiv:2004.00690}},
  \href {https://doi.org/10.1103/PhysRevC.102.034905}
  {\path{doi:10.1103/PhysRevC.102.034905}}.

\bibitem{Bernhard:2016tnd}
J.~E. Bernhard, J.~S. Moreland, S.~A. Bass, J.~Liu, U.~Heinz, {Applying
  Bayesian parameter estimation to relativistic heavy-ion collisions:
  simultaneous characterization of the initial state and quark-gluon plasma
  medium}, Phys. Rev. C 94~(2) (2016) 024907.
\newblock \href {http://arxiv.org/abs/1605.03954} {\path{arXiv:1605.03954}},
  \href {https://doi.org/10.1103/PhysRevC.94.024907}
  {\path{doi:10.1103/PhysRevC.94.024907}}.

\bibitem{Bass:2017jkj}
S.~A. Bass, J.~E. Bernhard, J.~S. Moreland, Determination of quark-gluon-plasma
  parameters from a global bayesian analysis, Nucl. Phys. A 967 (2017) 67--73.
\newblock \href {https://doi.org/10.1016/j.nuclphysa.2017.05.052}
  {\path{doi:10.1016/j.nuclphysa.2017.05.052}}.

\bibitem{Nijs:2020roc}
G.~Nijs, W.~van~der Schee, U.~G{\"u}rsoy, R.~Snellings, Bayesian analysis of
  heavy ion collisions with the heavy ion computational framework trajectum,
  Phys. Rev. C 103~(5) (2021) 054909.
\newblock \href {http://arxiv.org/abs/2010.15134} {\path{arXiv:2010.15134}},
  \href {https://doi.org/10.1103/PhysRevC.103.054909}
  {\path{doi:10.1103/PhysRevC.103.054909}}.

\bibitem{LeCam1986}
L.~Le~Cam, Asymptotic Methods in Statistical Decision Theory, Springer, New
  York, 1986.

\bibitem{Kass1990}
R.~E. Kass, L.~Tierney, J.~B. Kadane, The validity of posterior expansions
  based on laplace's method, in: S.~Geisser, J.~S. Hodges, S.~J. Press,
  A.~Zellner (Eds.), Bayesian and Likelihood Methods in Statistics and
  Econometrics, Elsevier, 1990, pp. 473--488.

\bibitem{Robert2007}
C.~Robert, Noninformative Prior Distributions, 2nd Edition, Springer, 2007, pp.
  127--141.

\bibitem{Huan:2022}
X.~Huan, et~al., Parameter estimation and uncertainty quantification using
  information geometry, Interface Focus 12 (2022) 20210054.

\bibitem{Lonardoni:2018nob}
D.~Lonardoni, S.~Gandolfi, J.~E. Lynn, C.~Petrie, J.~Carlson, K.~E. Schmidt,
  A.~Schwenk, {Auxiliary field diffusion Monte Carlo calculations of light and
  medium-mass nuclei with local chiral interactions}, Phys. Rev. C 97~(4)
  (2018) 044318.
\newblock \href {http://arxiv.org/abs/1802.08932} {\path{arXiv:1802.08932}},
  \href {https://doi.org/10.1103/PhysRevC.97.044318}
  {\path{doi:10.1103/PhysRevC.97.044318}}.

\bibitem{Frosini:2021ddm}
M.~Frosini, T.~Duguet, J.-P. Ebran, B.~Bally, H.~Hergert, T.~R. Rodr\'\i{}guez,
  R.~Roth, J.~Yao, V.~Som\`a, {Multi-reference many-body perturbation theory
  for nuclei: III. Ab initio calculations at second order in PGCM-PT}, Eur.
  Phys. J. A 58~(4) (2022) 64.
\newblock \href {http://arxiv.org/abs/2111.01461} {\path{arXiv:2111.01461}},
  \href {https://doi.org/10.1140/epja/s10050-022-00694-x}
  {\path{doi:10.1140/epja/s10050-022-00694-x}}.

\bibitem{Elhatisari:2017eno}
S.~Elhatisari, E.~Epelbaum, H.~Krebs, T.~A. L\"ahde, D.~Lee, N.~Li, B.-n. Lu,
  U.-G. Mei\ss{}ner, G.~Rupak, {Ab initio Calculations of the Isotopic
  Dependence of Nuclear Clustering}, Phys. Rev. Lett. 119~(22) (2017) 222505.
\newblock \href {http://arxiv.org/abs/1702.05177} {\path{arXiv:1702.05177}},
  \href {https://doi.org/10.1103/PhysRevLett.119.222505}
  {\path{doi:10.1103/PhysRevLett.119.222505}}.

\bibitem{Meissner:2014lgi}
U.-G. Mei\ss{}ner, {A new tool in nuclear physics: Nuclear lattice
  simulations}, Nucl. Phys. News. 24~(4) (2014) 11--15.
\newblock \href {http://arxiv.org/abs/1505.06997} {\path{arXiv:1505.06997}},
  \href {https://doi.org/10.1080/10619127.2014.972167}
  {\path{doi:10.1080/10619127.2014.972167}}.

\bibitem{Amari2000}
S.-i. Amari, H.~Nagaoka, Methods of Information Geometry, Vol. 191 of
  Translations of Mathematical Monographs, American Mathematical Society, 2000.

\bibitem{Kennedy2001}
M.~C. Kennedy, A.~O'Hagan, Bayesian calibration of computer models, Journal of
  the Royal Statistical Society: Series B (Statistical Methodology) 63~(3)
  (2001) 425--464.
\newblock \href {https://doi.org/10.1111/1467-9868.00294}
  {\path{doi:10.1111/1467-9868.00294}}.

\bibitem{Higdon2008}
D.~Higdon, J.~Gattiker, B.~Williams, W.~Rightley, Computer experiments with
  functional inputs and outputs, Journal of the American Statistical
  Association 103~(482) (2008) 570--583.
\newblock \href {https://doi.org/10.1198/016214507000000619}
  {\path{doi:10.1198/016214507000000619}}.

\bibitem{Bernhard2016}
J.~E. Bernhard, S.~Moreland, S.~A. Bass, Bayesian estimation of the specific
  shear and bulk viscosity of quark--gluon plasma, Nature Physics 15 (2019)
  1113--1117.
\newblock \href {https://doi.org/10.1038/s41567-019-0611-8}
  {\path{doi:10.1038/s41567-019-0611-8}}.

\bibitem{Moreland2020}
S.~Moreland, J.~Bernhard, S.~A. Bass, Estimating the impact parameter of
  heavy-ion collisions with bayesian inference, Physical Review C 101 (2020)
  024911.
\newblock \href {https://doi.org/10.1103/PhysRevC.101.024911}
  {\path{doi:10.1103/PhysRevC.101.024911}}.

\end{thebibliography}


\appendix

\section{Bayesian response-model analysis}
\label{app:bayesian}

The Bayesian analysis is based on a smooth response model~\cite{Amari2000} for the multiparticle correlation observables over the scanned Woods--Saxon parameter space. The response model provides a continuous representation of the simulated observables, allowing the likelihood to be evaluated between the discrete Monte Carlo configurations without generating additional events~\cite{Kennedy2001,Higdon2008,Bernhard2016,Moreland2020}. The uncertainty model combines the statistical uncertainties of the Monte Carlo calculations with an effective model-discrepancy uncertainty estimated from the residuals of the response model. The response analysis is performed separately for each centrality interval. We highlight the $0$--$1\%$ centrality class here in the Bayesian and information-geometric analysis because it provides the largest sensitivity to intrinsic nuclear deformation; the centrality dependence of the response is discussed in the main text.

\subsection{Two-dimensional response model}
\label{app:bayesian2D}

We first consider the two-dimensional WS parameter space
$(a_0,\beta_2)$, with octupole deformation $\beta_3$ set to zero. The scanned values of $a_0$ and $\beta_2$
define a discrete grid on which the multiparticle correlation observables are
calculated. To construct a numerically well-conditioned response model, the
physical parameters are mapped to scaled coordinates,
\begin{equation}\label{scale-variables}
	x_{a_0} =
	\frac{a_0-a_{0,\min}}
	{a_{0,\max}-a_{0,\min}},
	\qquad
	x_{\beta_2} =
	\frac{\beta_2-\beta_{2,\min}}
	{\beta_{2,\max}-\beta_{2,\min}} .
\end{equation}
This transformation leaves the physical parameter dependence unchanged while
improving the numerical conditioning of the polynomial fit and making the
magnitudes of the different expansion coefficients more comparable.

For each multiparticle correlation observable $X$, we represent the
parameter dependence by a two-dimensional polynomial expansion,
\begin{equation}
	X(x_{a_0},x_{\beta_2})
	=
	\sum_{i=0}^{N_{a_0}}
	\sum_{j=0}^{N_{\beta_2}}
	c_{ij}\,
	x_{a_0}^i x_{\beta_2}^j .
	\label{eq:response_model_2D}
\end{equation}
The expansion order is chosen such that the response model reproduces the
simulated parameter dependence while avoiding unnecessary flexibility. In
practice, the polynomial order is fixed consistently for all observables and
centrality intervals. The coefficients $c_{ij}$ are determined from the
Monte Carlo results by a weighted least-squares procedure, with the weights
set by the statistical uncertainties of the corresponding simulated
observables.

The fitted coefficients provide a compact representation of the complete
parameter dependence. In particular, the derivatives required for the local
sensitivity analysis are obtained analytically from the response model,
\begin{equation}
	\frac{\partial X}{\partial a_0}
	=
	\frac{1}{a_{0,\max}-a_{0,\min}}
	\sum_{i,j}
	i\,c_{ij}\,
	x_{a_0}^{i-1}x_{\beta_2}^{j},
	\label{eq:dOda_2D}
\end{equation}
and
\begin{equation}
	\frac{\partial X}{\partial\beta_2}
	=
	\frac{1}{\beta_{2,\max}-\beta_{2,\min}}
	\sum_{i,j}
	j\,c_{ij}\,
	x_{a_0}^{i}x_{\beta_2}^{j-1}.
	\label{eq:dOdb2_2D}
\end{equation}
Thus, the Jacobian, local sensitivity vectors, and local degeneracy directions
used in the main text are obtained directly from the continuous response
surface rather than from finite differences of individual Monte Carlo
points.

\begin{figure*}[t!]
	\hspace*{.5cm}\includegraphics[scale=.29]{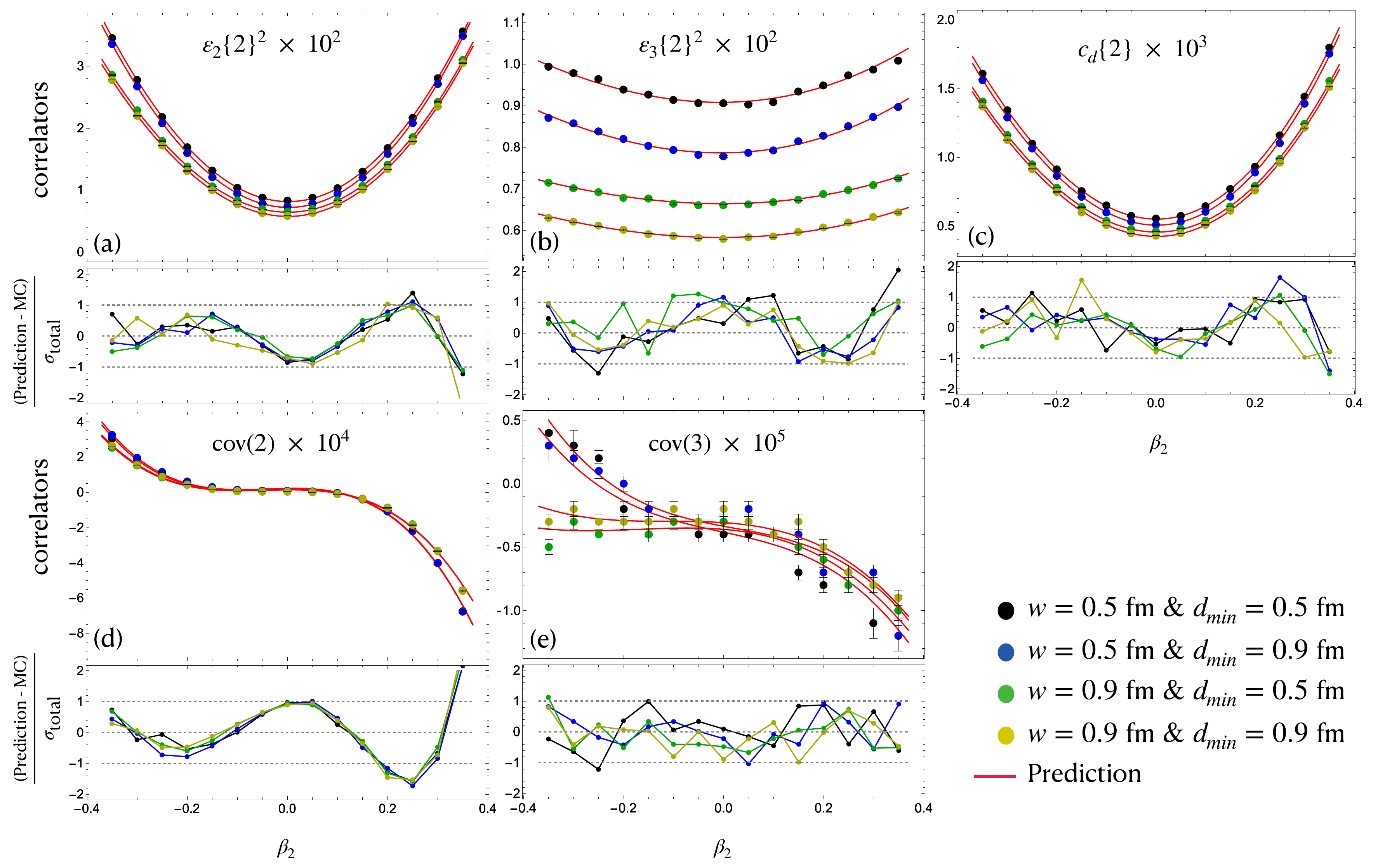}
	\begin{picture}(0,0)
		\put(-140,127){{\fontsize{11}{11}\selectfont \textcolor{black}{U+U @ 193 GeV}}}
		\put(-125,110){{\fontsize{11}{11}\selectfont \textcolor{black}{$a_0=0.55$ fm}}}
	\end{picture}	
	\caption{
		Validation of the two-dimensional Bayesian response model for
		$^{238}$U+$^{238}$U collisions at $\sqrt{s_{NN}}=193~\mathrm{GeV}$ and
		$0$--$1\%$ centrality. The upper panels compare the Monte Carlo
		calculations (symbols) with the corresponding response-model
		predictions (solid red curves) as a function of the quadrupole
		deformation $\beta_2$, at fixed $a_0=0.40~\mathrm{fm}$, for
		$\varepsilon_2\{2\}^2$, $\varepsilon_3\{2\}^2$, $c_d\{2\}$,
		$\mathrm{cov}(2)$, and $\mathrm{cov}(3)$. The four colors correspond
		to $(w,d_{\min})=(0.5,0.5)$, $(0.5,0.9)$, $(0.9,0.5)$, and
		$(0.9,0.9)~\mathrm{fm}$, respectively. The lower panels show the
		corresponding normalized residuals,
		$(O_{\rm pred}-O_{\rm MC})/\sigma_{\rm tot}$, where
		$\sigma_{\rm tot}$ denotes the total uncertainty used in the Bayesian
		response model. The horizontal dashed lines indicate
		$r_{\mathcal O}=0$ and $r_{\mathcal O}=\pm1$. The close agreement
		between the response model and the Monte Carlo calculations, together
		with the absence of systematic residuals, demonstrates that the
		polynomial response model provides an adequate representation of the
		scanned two-dimensional parameter dependence.}
	\label{fig:response-model}
\end{figure*}
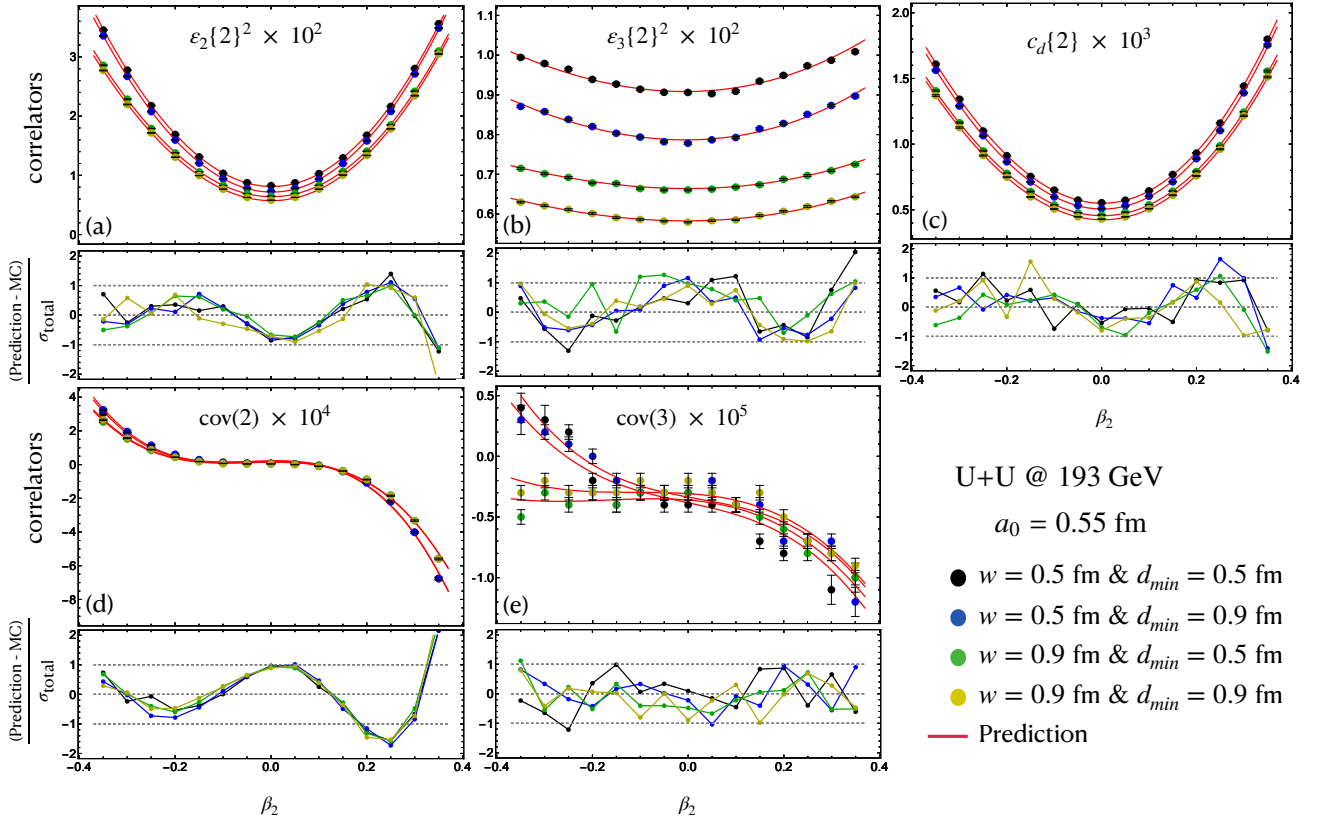

The uncertainty of the response model is determined from the covariance of
the fitted coefficients. Denoting the vector of fitted coefficients by
$\mathbf c$, we obtain its covariance matrix from the weighted regression,
\begin{equation}
	\mathbf C_c
	=
	\left(\mathbf Y^{T}\mathbf W\mathbf Y\right)^{-1},
	\label{eq:coefficient_covariance_2D}
\end{equation}
where $\mathbf Y$ is the design matrix constructed from the polynomial basis
and $\mathbf W$ contains the inverse variances of the simulated observables.
The resulting coefficient uncertainties are propagated to the response and
its derivatives. For a parameter point $\boldsymbol{x}$, the model variance
of the predicted observable is therefore
\begin{equation}
	\sigma_{\rm model}^2(\boldsymbol{x})
	=
	\boldsymbol{\phi}^{T}(\boldsymbol{x})
	\mathbf C_c
	\boldsymbol{\phi}(\boldsymbol{x}),
	\label{eq:model_variance_2D}
\end{equation}
where $\boldsymbol{\phi}$ denotes the vector of polynomial basis functions.
The same coefficient covariance is used to propagate uncertainties to the
local derivatives entering the sensitivity and information-geometry
analysis.
%

For the two-dimensional analysis, we work in the scaled WS parameter
coordinates
$\mathbf{x}=(x_{a_0},x_{\beta_2})$, defined above. At a chosen
reference configuration $\mathbf{x}_{\rm ref}$, the corresponding
correlation vector
$\mathbf{y}=\boldsymbol{\mu}(\mathbf{x}_{\rm ref})$
is used as the reference response. The likelihood for a trial parameter
point $\mathbf{x}$ is then constructed from the difference between this
reference response and the response-model prediction
$\boldsymbol{\mu}(\mathbf{x})$:
\begin{equation}
	\mathcal L(\mathbf{y}|\mathbf{x})
	\propto
	\exp\left[
	-\frac{1}{2}
	\left(\mathbf y-\boldsymbol{\mu}(\mathbf{x})\right)^T
	\mathbf\Sigma^{-1}
	\left(\mathbf y-\boldsymbol{\mu}(\mathbf{x})\right)
	\right],
	\label{eq:bayes_likelihood_2D}
\end{equation}
where $\mathbf\Sigma$ contains the combined statistical, response-model,
and model-discrepancy uncertainties. The posterior distribution is then
given by
\begin{equation}
	p(\mathbf{x}|\mathbf y)
	\propto
	\mathcal L(\mathbf y|\mathbf{x})p(\mathbf{x}),
	\label{eq:bayes_posterior_2D}
\end{equation}
with the prior restricted to the scanned physical parameter region
mapped onto the corresponding scaled coordinates.

The response-model construction is performed independently for each
multiparticle correlation observable. This allows the coefficient structure
to be inspected directly and makes it possible to identify which terms in
the polynomial expansion generate the dependence on $a_0$ and $\beta_2$.
In particular, the coefficients multiplying powers of $a_0$ quantify the
direct response to surface diffuseness, while mixed terms such as
$x_{a_0}^i x_{\beta_2}^j$ encode the coupling between diffuseness and quadrupole
deformation. These mixed contributions are especially important for
understanding the local deformation--diffuseness degeneracy discussed in the
main text.

\subsection{Response-model validation and uncertainty propagation}
\label{app:validation2D}

\begin{figure*}[t!]
	\hspace*{0cm}\includegraphics[scale=.34]{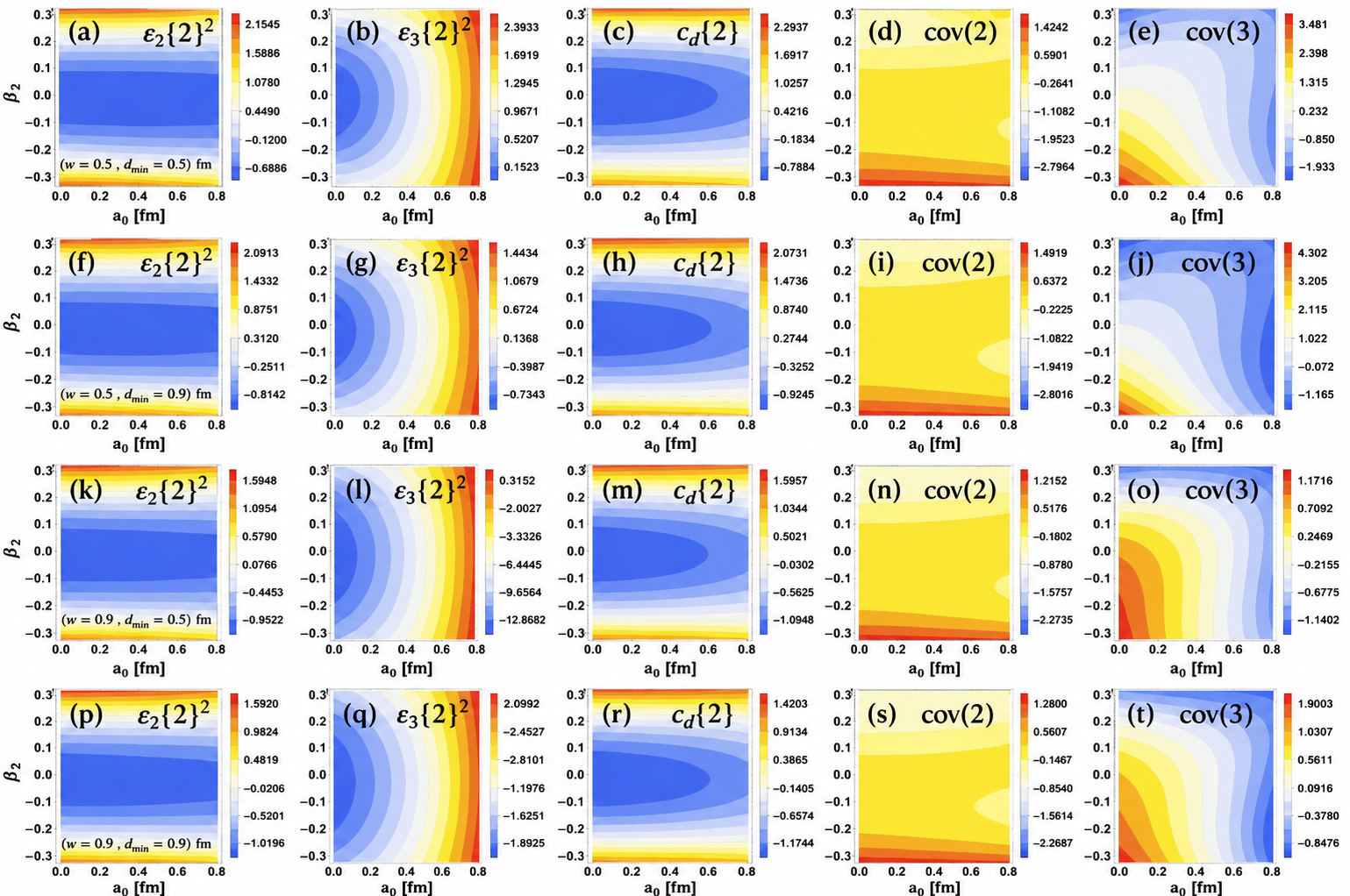}
	\caption{
		Posterior-mean response maps of the five initial-state observables
		in the $(a_0,\beta_2)$ plane.  From left to right, the columns show
		$\varepsilon_2\{2\}^2$, $\varepsilon_3\{2\}^2$, $c_d\{2\}$,
		$\mathrm{cov}(2)$, and $\mathrm{cov}(3)$.  From top to bottom, the
		rows correspond to $(w,d_{\min})=(0.5,0.5)$,
		$(0.5,0.9)$, $(0.9,0.5)$, and $(0.9,0.9)$~fm.  The maps are obtained
		by evaluating the posterior response model on a continuous grid
		in the WS parameter space.  They illustrate the
		characteristic response directions of the observables and their
		variation with the treatment of the nucleon width and
		short-range nucleon correlations.}
	\label{fig:app_posterior}
\end{figure*}

Before performing the Bayesian parameter inference, we validate the response
model against the underlying Monte Carlo calculations. For each observable,
the response model is fitted independently over the scanned
$(a_0,\beta_2)$ parameter space using the polynomial basis described above.
Figure~\ref{fig:response-model} compares the resulting predictions with
the Monte Carlo values for the four combinations of nucleon width and
minimum inter-nucleon separation considered in the analysis. The upper
panels show the observables as functions of $\beta_2$ at fixed
$a_0=0.55~\mathrm{fm}$, with the symbols denoting the Monte Carlo results
and the solid curves the corresponding response-model predictions. The
different colors correspond to the four $(w,d_{\min})$ configurations.

The response model reproduces the characteristic deformation dependence of
all observables over the full range of $\beta_2$. In particular, the
approximately parabolic behavior of the elliptic eccentricity and transverse
size fluctuation, as well as the non-monotonic structure of the covariance
observables, is well captured. The agreement remains good for the
three-particle observables, whose smaller absolute magnitudes make their
response more sensitive to the quality of the interpolation. This agreement
indicates that the polynomial representation provides a sufficiently smooth
description of the calculated response without washing out the relevant
deformation dependence.

The lower panels show the normalized residual for each correlator $X$, defined as 
\begin{equation}
	\frac{
		X_{\rm pred}(\beta_2)-X_{\rm MC}(\beta_2)
	}{
		\sigma_{\rm tot}(\beta_2)
	},
	\label{eq:normalized_response_residual}
\end{equation}
where $\sigma_{\rm tot}$ combines the Monte Carlo uncertainty with the
model-discrepancy scale used in the Bayesian analysis. The residuals remain
predominantly within the expected $\mathcal{O}(1)$ range for all four
initial-state configurations, with no systematic deviation across the
deformation interval. The larger fluctuations visible for the covariance
observables reflect their smaller absolute scale and stronger sensitivity to
the details of the initial geometry. Overall, Fig.~\ref{fig:response-model}
demonstrates that the response model provides an adequate representation of
the Monte Carlo calculations for subsequent Bayesian response-model analysis.

\begin{figure*}[t!]
	\hspace*{0cm}\includegraphics[scale=.34]{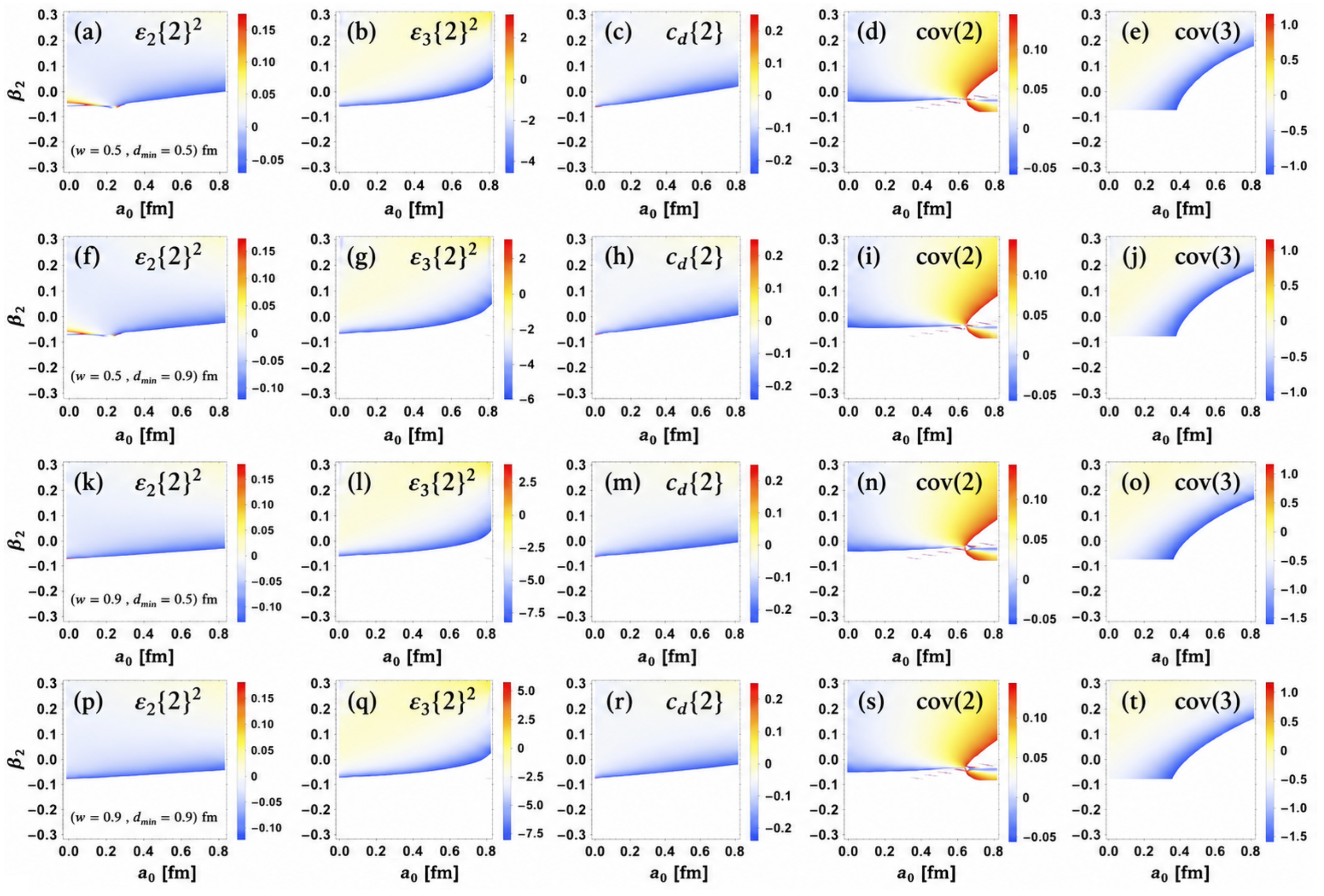}
	\caption{
		Local deformation--diffuseness degeneracy maps obtained from the
		posterior response model.  The panels follow the same ordering as
		Fig.~\ref{fig:app_posterior}: from left to right,
		$\varepsilon_2\{2\}^2$, $\varepsilon_3\{2\}^2$, $c_d\{2\}$,
		$\mathrm{cov}(2)$, and $\mathrm{cov}(3)$, while the four rows
		correspond to $(w,d_{\min})=(0.5,0.5)$, $(0.5,0.9)$,
		$(0.9,0.5)$, and $(0.9,0.9)$~fm.  The color scale represents the
		local constant-observable slope
		$d\beta_2/da_0$, obtained from the derivatives of the posterior
		response model.  Large or rapidly varying values indicate regions
		in which the response to $\beta_2$ becomes weak and the local
		deformation--diffuseness degeneracy is correspondingly enhanced.
		}
	\label{fig:app_degeneracy}
\end{figure*}
\subsection{Posterior constraints and local parameter degeneracies}
\label{app:posterior_degeneracy}

The Bayesian response model described above provides a continuous
representation of each initial-state observable over the
$(a_0,\beta_2)$ parameter space.  Rather than evaluating the response
only at the simulated WS configurations, the posterior distribution of
the polynomial coefficients allows the observable and its uncertainty
to be reconstructed at arbitrary points in the scanned parameter
region.  This provides a convenient way to visualize both the
parameter dependence of the observables and the regions in which the
response is poorly constrained.

Figure~\ref{fig:app_posterior} shows the posterior-mean response maps
for the five observables for the four combinations of nucleon width
and minimum nucleon separation.  The columns correspond to
$\varepsilon_2\{2\}^2$, $\varepsilon_3\{2\}^2$, $c_d\{2\}$,
$\mathrm{cov}(2)$, and $\mathrm{cov}(3)$, while the rows correspond to
$(w,d_{\min})=(0.5,0.5)$, $(0.5,0.9)$, $(0.9,0.5)$, and
$(0.9,0.9)$~fm.  Several common features are evident across the four
initial-state configurations.  The elliptic observables exhibit a
strong dependence on $\beta_2$, with comparatively weak variation
along the $a_0$ direction.  In contrast, $\varepsilon_3\{2\}^2$
shows a pronounced variation with $a_0$, while its dependence on
$\beta_2$ is qualitatively different from that of the elliptic
observables.  The transverse-size correlator $c_d\{2\}$ displays a
response intermediate between these two behaviors.  The covariance
observables exhibit more complicated structures, including regions
where their response changes rapidly with either nuclear parameter.

The persistence of these structures when $w$ and $d_{\min}$ are
varied indicates that the characteristic response directions are not
generated solely by one particular choice of short-range nucleon
correlations.  At the same time, the quantitative changes between the
four rows demonstrate that the initial-state prescription affects the
strength of the response.  The posterior response maps therefore
provide the basis for separating robust geometric features from
model-dependent details.

The local parameter degeneracy can be obtained directly from the
gradient of the posterior response defined in Eqs.~\ref{satis} and \ref{deg}. 
Figure~\ref{fig:app_degeneracy} displays this quantity over the same
parameter space and for the same four initial-state configurations.
The resulting maps make the observable dependence of the
deformation--diffuseness degeneracy explicit.  For
$\varepsilon_2\{2\}^2$ and $c_d\{2\}$, the local degeneracy direction
is generally close to the $a_0$ axis, reflecting the comparatively
strong response to $\beta_2$.  In contrast, the corresponding maps
for $\varepsilon_3\{2\}^2$ exhibit substantially larger slopes,
indicating that variations of $a_0$ can be compensated by appreciable
changes in $\beta_2$ while leaving the observable approximately
unchanged.  The covariance observables show more localized and
rapidly varying degeneracy directions, particularly in regions where
the derivative with respect to $\beta_2$ becomes small.  Such regions
naturally lead to large values of Eq.~(\ref{deg}) and
should therefore not be interpreted as uniformly enhanced
sensitivity to $a_0$.

An important feature of Fig.~\ref{fig:app_degeneracy} is that the
degeneracy direction is not constant throughout the parameter space.
Its magnitude and, for some observables, its sign change as
$(a_0,\beta_2)$ is varied.  The ambiguity between surface diffuseness
and deformation is therefore intrinsically local rather than a single
global direction in parameter space.  This provides a direct
connection between the response maps and the Fisher-information
analysis used in the main text: the weakly constrained directions
arise because different nuclear-parameter variations can produce
nearly equivalent changes in the measured correlations.


\begin{figure*}[t!]
	\hspace*{0cm}\includegraphics[scale=.27]{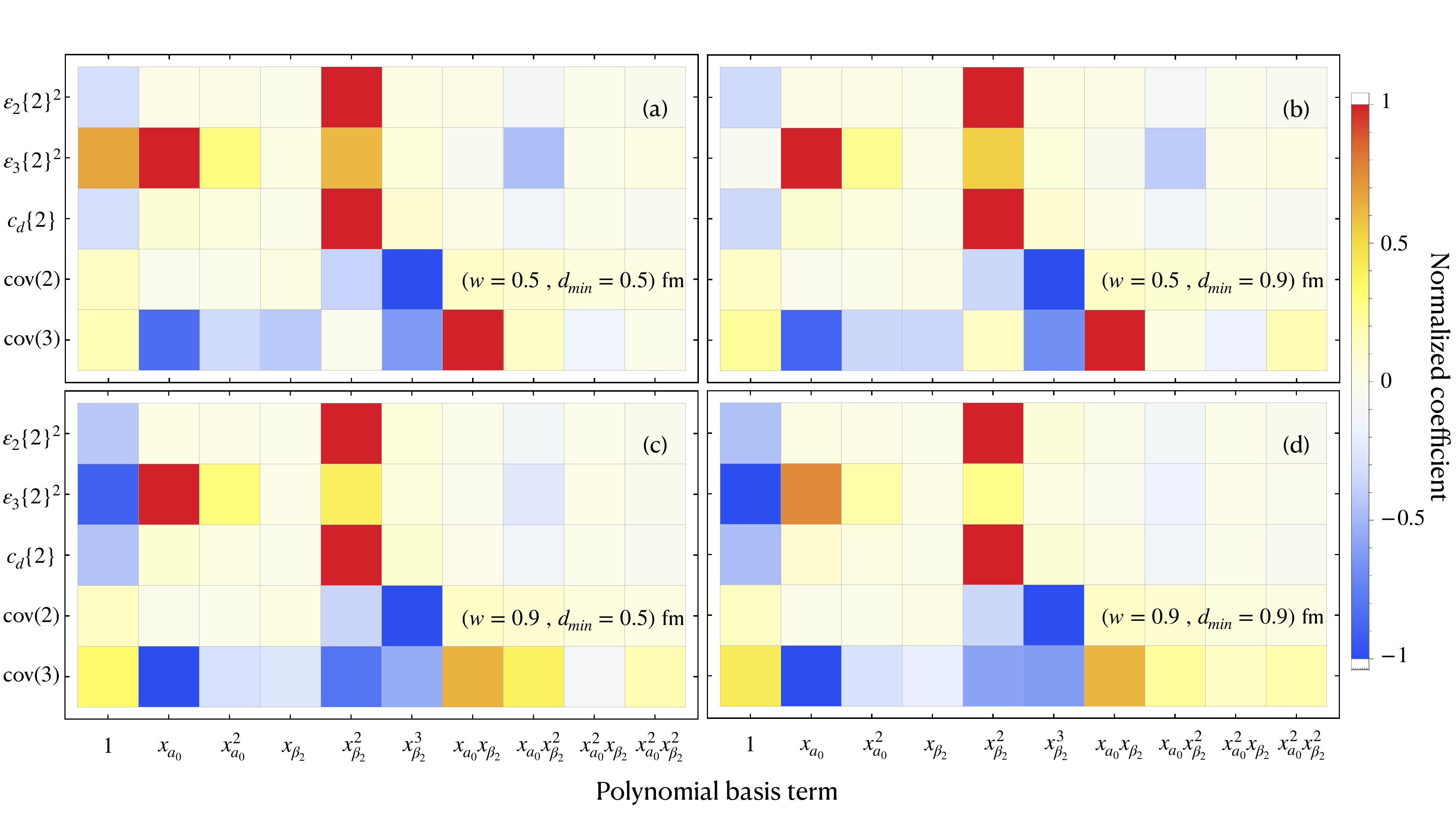}
	\caption{
		Decomposition of the two-dimensional Bayesian response model into
		its polynomial basis coefficients.  Each panel corresponds to one
		initial-state configuration:
		(a) $(w,d_{\min})=(0.5,0.5)$~fm,
		(b) $(0.5,0.9)$~fm,
		(c) $(0.9,0.5)$~fm, and
		(d) $(0.9,0.9)$~fm.  The rows correspond to
		$\varepsilon_2\{2\}^2$, $\varepsilon_3\{2\}^2$, $c_d\{2\}$,
		$\mathrm{cov}(2)$, and $\mathrm{cov}(3)$, while the columns show
		the constant, pure-$a_0$, pure-$\beta_2$, and mixed polynomial
		terms of Eq.~(\ref{eq:app_polynomial}).  The color scale denotes
		the normalized posterior coefficient.  The decomposition
		illustrates the relative importance of linear, nonlinear, and
		mixed parameter dependences in the fitted response model.  The
		coefficient values themselves should not be interpreted as
		pointwise contributions, since the corresponding basis functions
		depend on the location in the $(a_0,\beta_2)$ parameter space.}
	\label{fig:app_decomposition}
\end{figure*}
\subsection{Polynomial decomposition of the local response}
\label{app:polynomial_decomposition}

The response model also provides a useful decomposition of the
observable dependence into contributions associated with the
individual nuclear parameters and their nonlinear couplings.  Using
the scaled variables introduced in Eq.~\ref{scale-variables},
the truncation of two-dimensional response model defined in Eq.~\ref{eq:response_model_2D} is written as
\begin{align}
	X(x_{a_0},x_\beta)
	=
	c_0
	&+c_{a_0} x_{a_0}+c_{a^2}x_{a_0}^2
	+c_\beta x_\beta+c_{\beta^2}x_\beta^2
	+c_{\beta^3}x_\beta^3
	+c_{a\beta}x_{a_0}x_\beta\nonumber\\&
	+c_{a\beta^2}x_{a_0}x_\beta^2
	+c_{a^2\beta}x_{a_0}^2x_\beta
	+c_{a^2\beta^2}x_{a_0}^2x_\beta^2 .
	\label{eq:app_polynomial}
\end{align}
We note that this form separates the leading dependence on the two nuclear
parameters from their nonlinear and mixed responses.  The coefficients
are obtained simultaneously from the Bayesian fit and therefore
include the information from the complete scanned parameter space.
Because the observables are standardized before the joint response
analysis, the coefficients provide a useful relative measure of the
importance of the corresponding basis functions.

Figure~\ref{fig:app_decomposition} summarizes the fitted coefficients
for all five observables and all four initial-state configurations.
Each panel corresponds to one choice of $(w,d_{\min})$, while the
rows represent the observables and the columns represent the ten
basis functions appearing in Eq.~(\ref{eq:app_polynomial}).  The
color scale shows the normalized posterior coefficient.

Several features of the decomposition clarify the response patterns
observed in Figs.~\ref{fig:app_posterior} and
\ref{fig:app_degeneracy}.  The coefficients multiplying the
quadrupole-dependent terms are particularly important for
$\varepsilon_2\{2\}^2$ and $c_d\{2\}$, consistent with the strong
deformation dependence of these observables.  In contrast,
$\varepsilon_3\{2\}^2$ contains a comparatively stronger dependence
on the $a_0$-related terms and on mixed combinations involving
$a_0$ and $\beta_2$.  The covariance observables contain a richer
mixture of linear, nonlinear, and mixed terms, reflecting their more
complicated response surfaces.

The mixed coefficients are especially relevant for the local
deformation--diffuseness degeneracy.  Terms such as
$x_{a_0} x_\beta$, $x_{a_0} x_\beta^2$, and $x_{a_0}^2x_\beta$ explicitly couple
the response to the two nuclear parameters and allow a change in
$a_0$ to be compensated locally by a change in $\beta_2$.  Their
presence demonstrates that the degeneracy is not simply a consequence
of two independent sensitivities to $a_0$ and $\beta_2$, but also
arises from nonlinear coupling between the two parameters.

The comparison among the four panels further shows that the dominant
coefficient pattern is relatively stable against changes in
$(w,d_{\min})$, although the magnitude of individual terms can vary.
This provides a complementary interpretation of the robustness
observed in the response maps.  In particular, the persistence of the
quadrupole-dominated structure of the elliptic observables and the
distinct $a_0$ dependence of the triangular observables can be traced
to corresponding patterns in the polynomial response coefficients.

It is important to emphasize that the coefficient magnitude alone is
not identical to the contribution of a term at a particular point in
parameter space.  The actual contribution of, for example, the
$x_{a_0} x_\beta$ term is $c_{a\beta}x_{a_0}x_\beta$ and therefore depends on
the location in the $(a_0,\beta_2)$ plane.  Figure~\ref{fig:app_decomposition}
should consequently be interpreted as a decomposition of the fitted
response model rather than as a point-by-point ranking of the
observable contributions.  Together with the posterior response and
local-degeneracy maps, it identifies which polynomial structures are
responsible for the different sensitivity directions.

\subsection{Extension to three-dimensional nuclear deformation}
\label{app:bayesian3D}

The two-dimensional analysis in $(a_0,\beta_2)$ can be naturally extended by including the octupole deformation parameter $\beta_3$. The nuclear
parameter vector is therefore
\begin{equation}
	\boldsymbol{\theta}
	=
	(a_0,\beta_2,\beta_3).
\end{equation}
The three-dimensional scan contains
\begin{equation}
	N_{\rm cfg}
	=
	12(a_0)\times17(\beta_2)\times8(\beta_3)
	=
	1632
\end{equation}
WS configurations, with
$a_0$ varied over the range $0$--$0.8$~fm, $\beta_2$ over
$-0.35$--$0.35$, and $\beta_3$ over $0$--$0.35$. As in the
two-dimensional analysis, the $0$--$1\%$ centrality class is used because
it provides the strongest sensitivity to intrinsic nuclear deformation.

The octupole parameter is restricted to $\beta_3\geq0$. For the
observables considered here, the random orientation of the nuclei
removes sensitivity to the sign of the intrinsic octupole deformation,
so that the positive branch contains the relevant information without
duplicating equivalent configurations.

The three-dimensional response is evaluated using the same five
multiparticle correlations,
\begin{equation}
	\mathbf X =
	\left(
	\varepsilon_2\{2\}^2,
	\varepsilon_3\{2\}^2,
	c_d\{2\},
	\mathrm{cov}(2),
	\mathrm{cov}(3)
	\right),
\end{equation}
and the response-model and likelihood construction described above.
The reference configuration is
\begin{equation}
	\boldsymbol{\theta}_{\rm ref}
	=
	(0.55,0.28,0),
\end{equation}
while the finite-octupole configuration
\begin{equation}
	\boldsymbol{\theta}_{\rm ref}
	=
	(0.55,0.28,0.1)
\end{equation}
is used to examine how the local constraints change once a nonzero
octupole deformation is present.

The important consequence of this extension is that the local
deformation--diffuseness degeneracy is no longer represented by a
single direction in the $(a_0,\beta_2)$ plane. Instead, the allowed
variations form a surface in the three-dimensional
$(a_0,\beta_2,\beta_3)$ space. Consequently, the relevant question is
not whether one particular parameter is constrained independently, but
which combinations of the three nuclear parameters are constrained by
the complete set of multiparticle correlations.

\subsection{Three-dimensional response decomposition}
\label{app:terms3D}

The inclusion of $\beta_3$ also extends the polynomial decomposition of
the local response. Using the scaled variables introduced above, each
observable is represented by the same 17-term three-dimensional basis,
\begin{equation}
	X(\boldsymbol{\theta})
	\simeq
	\sum_k c_k B_k(x_{a_0},x_2,x_3),
\end{equation}
with
\begin{align}
	B_k=\{&
	1,\,
	x_{a_0},\,
	x_{a_0}^2,\,
	x_{a_0}^3,\,
	x_2^2,\,
	x_2^3,\,
	x_3^2,\,
	x_3^3,
	\nonumber\\
	&
	x_{a_0} x_2^2,\,
	x_{a_0}^2x_2,\,
	x_{a_0}^2x_2^2,\,
	x_{a_0} x_3^2,\,
	x_{a_0}^2x_3,
	\nonumber\\
	&
	x_{a_0}^2x_3^2,\,
	x_2x_3,\,
	x_2^2x_3,\,
	x_2x_3^2
	\}.
\end{align}
The corresponding design matrix contains $1632$ configurations and has
dimensions $X\in\mathbb{R}^{1632\times17}$. Its rank and condition
number are
\begin{equation}
	{\rm rank}(X)=17,
	\qquad
	\kappa(X)
	=
	\frac{\sigma_{\max}(X)}
	{\sigma_{\min}(X)}
	=
	6.29\times10^{1},
\end{equation}
where $\sigma_{\max}(X)$ and $\sigma_{\min}(X)$ denote the largest and
smallest singular values of the design matrix, respectively. Full column rank ensures that the $17$ basis coefficients are independently
identifiable within the chosen polynomial basis, while the condition
number quantifies the degree of numerical ill-conditioning of the
response decomposition. The value obtained here indicates that the
basis is sufficiently well conditioned for the present analysis.

More directly, the relative importance of the three nuclear directions
can be seen from the local Jacobian. At the reference point
$(a_0,\beta_2,\beta_3)=(0.55,0.28,0)$, the response is
\begin{equation}
	J=
	\begin{pmatrix}
		-3.47\times10^{-3} & 1.18\times10^{-1} & -5.90\times10^{-3}\\
		3.70\times10^{-3} & 6.32\times10^{-3} & -5.59\times10^{-3}\\
		4.16\times10^{-5} & 4.94\times10^{-3} & -1.90\times10^{-4}\\
		3.66\times10^{-5} &-2.43\times10^{-3} & -2.17\times10^{-4}\\
		-2.27\times10^{-5} &-2.69\times10^{-5} & -1.22\times10^{-4}
	\end{pmatrix},
\end{equation}
where the columns correspond to
$(a_0,\beta_2,\beta_3)$ and the rows to the five observables in the
order defined above.

The quadrupole response remains particularly strong for
$\varepsilon_2\{2\}$, with
\begin{equation}
	\left.
	\frac{\partial\varepsilon_2\{2\}}
	{\partial\beta_2}
	\right|_{\rm ref}
	=
	1.18\times10^{-1},
\end{equation}
whereas the response to $a_0$ is substantially smaller. In contrast,
the sensitivity of $\varepsilon_3\{2\}$ to $\beta_3$ becomes much more
pronounced when a finite octupole deformation is introduced. At
$(0.55,0.28,0.1)$,
\begin{equation}
	\left.
	\frac{\partial\varepsilon_3\{2\}}
	{\partial\beta_3}
	\right|_{\rm ref}
	=
	3.48\times10^{-2},
\end{equation}
compared with
\begin{equation}
	\left.
	\frac{\partial\varepsilon_3\{2\}}
	{\partial\beta_3}
	\right|_{\beta_3=0}
	=
	-5.59\times10^{-3}.
\end{equation}
Thus, the response decomposition shows explicitly that the information
carried by $\varepsilon_3\{2\}$ changes with the underlying nuclear
configuration: a finite octupole deformation activates a substantially
stronger response in the $\beta_3$ direction.

The remaining observables provide additional, although weaker,
derivatives with respect to the nuclear parameters. In particular,
$c_d\{2\}$ and $\mathrm{cov}(2)$ retain a comparatively strong
dependence on $\beta_2$, while $\mathrm{cov}(3)$ develops a stronger
$\beta_3$ response at finite octupole deformation. The numerical
decomposition therefore demonstrates that the five correlations do not
respond to identical combinations of nuclear parameters, providing the
complementarity required for a multidimensional constraint.

\subsection{Three-dimensional information geometry}
\label{app:geometry3D}

The multidimensional response is most conveniently summarized through
the singular values of the uncertainty-weighted Jacobian introduced in
Eq.~\ref{eq:svd}. As mentioned, these singular values quantify the strength of the independently
constrained combinations of $(a_0,\beta_2,\beta_3)$, while the
corresponding right-singular vectors specify their directions in nuclear
parameter space.

At the reference configuration $(a_0,\beta_2,\beta_3)=(0.55,0.28,0)$,
the singular values are
\begin{equation}
	(S_1,S_2,S_3)
	=
	(671.2,\;39.2,\;16.0).
\end{equation}
Thus,
\begin{equation}
	\frac{S_2}{S_1}\simeq0.058,
	\qquad
	\frac{S_3}{S_1}\simeq0.024.
\end{equation}
The hierarchy indicates that one parameter combination is strongly
constrained, while the second and, in particular, the third direction
contain substantially less information. The weakest direction is
\begin{equation}
	\mathbf v_3
	=
	(0.865,\;0.024,\;0.501),
\end{equation}
showing that the locally least-constrained combination is dominated by
$a_0$ with a substantial $\beta_3$ component and only a small
$\beta_2$ contribution.

The information geometry changes significantly when the nuclear
configuration is moved to $(a_0,\beta_2,\beta_3)=(0.55,0.28,0.1)$.
The singular values become
\begin{equation}
	(S_1,S_2,S_3)
	=
	(655.4,\;202.5,\;13.0),
\end{equation}
giving
\begin{equation}
	\frac{S_2}{S_1}\simeq0.309,
	\qquad
	\frac{S_3}{S_1}\simeq0.020.
\end{equation}
Hence, the second independently constrained direction becomes much
stronger once a finite octupole deformation is present, whereas the
weakest direction remains poorly constrained. The corresponding weak
direction is
\begin{equation}
	\mathbf v_3
	=
	(0.996,\;0.008,\;-0.093),
\end{equation}
which is now almost aligned with the $a_0$ direction.

The configuration dependence is particularly clear from the singular
values obtained at representative deformation points:
\begin{center}
	\begin{tabular}{c|ccc}
		$(\beta_2,\beta_3)$
		& $S_1$ & $S_2$ & $S_3$\\
		\hline
		$(0.28,0)$   & 671.2 & 39.2  & 16.0\\
		$(0.28,0.1)$ & 655.4 & 202.5 & 13.0\\
		$(0.28,0.3)$ & 684.5 & 557.5 & 13.9\\
		$(0.16,0.3)$ & 584.7 & 369.7 & 16.1\\
		$(0,0.3)$    & 538.9 & 66.8  & 27.5\\
		$(0,0.1)$    & 208.2 & 30.0  & 8.93
	\end{tabular}
\end{center}

The ratio $S_2/S_1$ reaches approximately $0.82$ at
$(\beta_2,\beta_3)=(0.28,0.3)$, demonstrating that two independent
parameter combinations can become comparably well constrained in a
finite-octupole configuration. In contrast, $S_3/S_1$ remains small,
approximately between $0.02$ and $0.06$ over the configurations
considered here. The persistence of this small singular value shows
that adding $\beta_3$ does not eliminate the local degeneracy; rather,
it changes the orientation and strength of the locally unconstrained
combination.

The evolution of the weakest direction provides an even more direct
description of this effect. For the sequence
\begin{align}
	&(0.28,0),\quad
	(0.28,0.1),\quad
	(0.28,0.3),\nonumber\\&
	(0.16,0.3),\quad
	(0,0.3),\quad
	(0,0.1),
\end{align}
the corresponding weak directions are
\begin{align}
	(0.28,0):&
	\quad
	(0.865,\;0.024,\;0.501),
	\nonumber\\
	(0.28,0.1):&
	\quad
	(0.996,\;0.008,\;-0.093),
	\nonumber\\
	(0.28,0.3):&
	\quad
	(1.000,\;0.012,\;0.001),
	\nonumber\\
	(0.16,0.3):&
	\quad
	(0.999,\;-0.034,\;-0.001),
	\nonumber\\
	(0,0.3):&
	\quad
	(-0.990,\;0.143,\;-0.001),
	\nonumber\\
	(0,0.1):&
	\quad
	(0.055,\;-0.998,\;0.012).
\end{align}
The weakest direction therefore evolves from a mixed
$a_0$--$\beta_3$ combination near $(\beta_2,\beta_3)=(0.28,0)$ to an
almost purely $a_0$ direction at finite octupole deformation. When the
quadrupole deformation is reduced, the weak direction can instead
rotate toward $\beta_2$. This demonstrates that the deformation--
diffuseness degeneracy is intrinsically local: there is no single
global direction in the three-dimensional WS parameter space
that remains unconstrained throughout the physically relevant region.

The same conclusion follows from the Fisher-information eigenvalues.
At $(0.28,0)$ they are
\begin{equation}
	\lambda_i
	=
	\{255.96,\;1537.23,\;450450.79\},
\end{equation}
while at $(0.28,0.1)$ they become
\begin{equation}
	\lambda_i
	=
	\{169.04,\;41013.58,\;429545.82\}.
\end{equation}
We characterize the hierarchy of local parameter constraints by the Fisher condition number,
\begin{equation}
	\kappa_F
	=
	\frac{\lambda_{\max}(F)}
	{\lambda_{\min}(F)}
	=
	\left(\frac{S_1}{S_3}\right)^2.
	\label{eq:fisher_condition}
\end{equation}
For the two configurations considered above, this gives
\begin{equation}
	\kappa_F\simeq1.76\times10^3
\end{equation}
and
\begin{equation}
	\kappa_F\simeq2.54\times10^3,
\end{equation}
respectively. These large values quantify the strong hierarchy between
the best- and worst-constrained local combinations of nuclear
parameters.

Overall, the three-dimensional analysis shows that the inclusion of
$\beta_3$ does not simply add a third independently measurable nuclear
parameter. Instead, it changes the local information geometry by
redistributing sensitivity among the three deformation directions.
Finite octupole deformation can substantially strengthen a second
independent parameter combination, while a weak direction persists and
rotates across the parameter space. The resulting picture is therefore
one of a configuration-dependent local degeneracy rather than a single
universal degeneracy between surface diffuseness and nuclear
deformation.

\end{document}